\documentclass[12pt,preprint]{aastex}

\newcommand{\kms}{\ensuremath{\rm km \, s^{-1}}}
\newcommand{\tauone}{\ensuremath{\tau_1}}
\newcommand{\tauex}{\ensuremath{\tau_1^{\rm loc}}}
\newcommand{\taugal}{\ensuremath{\tau_1^{\rm gal}}}
\newcommand{\tautot}{\ensuremath{\tau_1^{\rm tot}}}
\newcommand{\kp}{\ensuremath{K^\prime}}
\newcommand{\ebv}{\ensuremath{E(B-V)}}
\newcommand{\ha}{\ensuremath{\rm H\alpha}}
\newcommand{\hb}{\ensuremath{\rm H\beta}}
\newcommand{\hg}{\ensuremath{\rm H\gamma}}
\newcommand{\hab}{\ensuremath{\ha/\hb}}
\newcommand{\fab}{\ensuremath{f_{\ha}/f_{\hb}}}
\newcommand{\ox}{\ensuremath{\log{\rm O/H}}}
\newcommand{\nh}{\ensuremath{n_{\rm H}}}
\newcommand{\nhi}{\ensuremath{N_{\rm HI}}}

\newcommand{\nhmol}{\ensuremath{N_{\rm H_2}}}
\newcommand{\hmol}{\ensuremath{\rm H_2}}
\newcommand{\ico}{\ensuremath{I_{\rm CO}}}
\newcommand{\vflat}{\ensuremath{V_{\rm flat}}}

\newcommand{\kext}{\ensuremath{k\_m\_ext}}
\newcommand{\ho}{\ensuremath{\rm H_0}}
\newcommand{\oii}{\ensuremath{{\rm [O II]} \lambda3727}}
\newcommand{\oiiia}{\ensuremath{{\rm [O III]} \lambda4959}}
\newcommand{\oiiib}{\ensuremath{{\rm [O III]} \lambda5007}}

\newcommand{\tautotmafii}{2.390}
\newcommand{\dtautotmafii}{0.203}
\newcommand{\tautotic}{0.866}
\newcommand{\dtautotic}{0.041}
\newcommand{\taumafii}{2.017}
\newcommand{\dtaumafii}{0.211}
\newcommand{\taucic}{0.692}
\newcommand{\dtaucic}{0.066}
\newcommand{\tausfdic}{0.639}
\newcommand{\dtausfdic}{0.102}
\newcommand{\tauic}{0.677}
\newcommand{\dtauic}{0.056}
\newcommand{\avmafii}{5.58}
\newcommand{\davmafii}{0.58}
\newcommand{\avic}{1.92}
\newcommand{\davic}{0.16}
\newcommand{\mtau}{\ensuremath{0.0057 \pm 0.0029}}
\newcommand{\btau}{\ensuremath{0.163 \pm 0.041}}
\newcommand{\mtfi}{\ensuremath{-10.52 \pm 0.49}}
\newcommand{\btfi}{\ensuremath{4.77 \pm 1.22}}
\newcommand{\rmstau}{0.17}
\newcommand{\rmsdnh}{0.16}
\newcommand{\rmstfi}{0.33}

\newcommand{\rmstfkext}{0.52}
\newcommand{\avedtauex}{0.10}

\newcommand{\avenh}{12.50}
\newcommand{\avednhi}{15}
\newcommand{\avedico}{26}
\newcommand{\avedfab}{11}
\newcommand{\muuma}{31.28}
\newcommand{\dmuuma}{0.08}
\newcommand{\mucoma}{34.45}
\newcommand{\dmucoma}{0.10}
\newcommand{\delzp}{0.236}
\newcommand{\hub}{\ensuremath{80.0 \, \kms \, \rm Mpc^{-1}}}
\newcommand{\spreaduma}{0.21}
\newcommand{\dismafii}{3.34}
\newcommand{\ddismafii}{0.56}
\newcommand{\dismafi}{2.85}
\newcommand{\ddismafi}{0.36}
\newcommand{\mumafii}{27.62}
\newcommand{\dmumafii}{0.36}
\newcommand{\muic}{27.41}
\newcommand{\dmuic}{0.12}
\newcommand{\disic}{3.03}
\newcommand{\ddisic}{0.17}
\newcommand{\disgrp}{3.03}
\newcommand{\ddisgrp}{0.15}
\newcommand{\depgrp}{0.49}
\newcommand{\spreadgrp}{0.53}
\newcommand{\dishubgrp}{3.17}
\newcommand{\spreadmi}{0.02}
\newcommand{\vlggrp}{253}

\slugcomment{Submitted to The Astrophysical Journal}
\shortauthors{Fingerhut et al.}
\shorttitle{Maffei 2}

\begin{document}

\title{
The Extinction and Distance of Maffei~2 and a New View of the IC~342/Maffei Group
}

\author{
Robin L. Fingerhut\altaffilmark{1},
Henry Lee\altaffilmark{1,2,3},
Marshall L. McCall\altaffilmark{1}, 
and Michael G. Richer\altaffilmark{2,4}
}

\altaffiltext{1}{
York University,
Department of Physics and Astronomy,
4700 Keele Street, Toronto, Ontario, Canada, M3J~1P3.
Email: {\tt rfinger@yorku.ca, mccall@yorku.ca}
}

\altaffiltext{2}{Visiting Astronomer, Kitt Peak National Observatory, 
National Optical Astronomy Observatories, which is operated by the
Association of Universities for Research in Astronomy, Inc. (AURA)
under cooperative agreement with the National Science Foundation.}

\altaffiltext{3}{Present location: 
Gemini Observatory, Southern Operations Center,
c/o AURA, Casilla 603, La Serena, Chile.
E-mail: {\tt hlee@gemini.edu}
}

\altaffiltext{4}{Observatorio Astron\'omico Nacional, P.O. Box 439027
San Diego, CA, 92143--9027 USA.
E-mail: {\tt richer@astrosen.unam.mx}
}

\begin{abstract}
We have obtained spectra of HII regions in the heavily obscured
spiral galaxy Maffei~2.
The observations have allowed for a determination of the Galactic extinction
of this galaxy
using a correlation between extinction and hydrogen column density
observed among spirals.
The technique reveals that the optical depth of Galactic dust at 1~$\mu$m
obscuring Maffei~2 is $\tauone=\taumafii \pm \dtaumafii$,
which implies that $A_V=\avmafii \pm \davmafii$~mag,
significantly higher than observed for the giant elliptical
Maffei~1 despite its similar latitude.
For comparison, we apply the same technique to IC~342,
a neighbouring spiral to Maffei~2 but with more moderate
obscuration by Galactic dust, owing to its higher Galactic latitude.
For this galaxy, we obtain $\tauone=\taucic \pm \dtaucic$,
which agrees within errors with the value of $\tausfdic \pm \dtausfdic$
derived from the reddening estimate of \citet{sfd98}.
We therefore adopt the weighted mean of $\tauone=\tauic \pm \dtauic$
for the extinction of IC~342, which implies that
$A_V=\avic \pm \davic$~mag.
A new distance estimate for Maffei~2 of $\dismafii \pm \ddismafii$~Mpc
is obtained from a self-consistent Tully-Fisher relation in $I$
adjusted to the NGC~4258 maser zero-point.
With our new measurement of $M_I$, Maffei~2
joins Maffei~1 and IC~342 as one of three
giant members of the
nearby IC~342/Maffei group of galaxies.
We present the revised properties of all three galaxies
based on the most accurate extinction and distance estimates
to date,
accounting for shifts
in the effective wavelengths of broadband filters
as this effect can be significant for highly reddened galaxies.
The revised distances are consistent with what would be
suspected for the Hubble Flow,
making it highly unlikely that the galaxies interacted
with the Local Group since the Big Bang.
\end{abstract}


\section{Introduction}

The spiral galaxy Maffei~2
was first detected by Paolo Maffei in 1968
on a near-infrared Schmidt plate \citep{maf68}.
Recent $I$-band photometry of this galaxy
reveals a large highly-inclined barred spiral of Hubble
type Sbc with isophotes visible out to $12^\prime$ from the
nucleus \citep{bm99}. If the galaxy lies within 
2\--6 Mpc as suggested in previous studies
\citep{bot71,spi73}, it must be among the dominant
galaxies in our Galactic neighbourhood,
which raises the question
of its dynamical role in the early evolution of
the Local Group.
Unfortunately, a precise distance to Maffei~2 has remained
elusive owing to the heavy obscuration by dust in the Milky Way disk
associated with its low Galactic latitude ($\delta=-0.33^\circ$).
Attempts to determine the Galactic extinction have
so far produced ambiguous results:
\citet{spi73} derived a value of 
$A_V = 6.3$~mag by comparing the
nuclear spectrum with that of M31,
and estimated that the galaxy suffers more
extinction than the giant elliptical Maffei~1 by about 1 magnitude in $V$.
However, a modern estimate of the reddening of Maffei~1
has been measured by Fingerhut et al. (2003; hereafter \citealt{fmd03})
using a well-defined
correlation between the $\rm Mg_2$ index and $V-I$ colour.
They find $A_V=4.68 \pm 0.18$~mag, which is over
1.5~mag lower than the result reported by \citet{spi73}.
There is clearly a need to revisit the problem
of the Galactic extinction of Maffei~2 before an attempt
can be made to determine its distance.

In this study, we derive the extragalactic extinction of Maffei~2
using a relationship among spiral galaxies
between the extragalactic extinction of
HII regions and the column density of extragalactic hydrogen gas along
the line of sight of the HII regions.
The correlation arises because the bulk of the extragalactic
extinction of an HII region is due to dust outside the region
of emitting gas, where the extinguishing material is
widely distributed and well-mixed with hydrogen gas
\citep{mrs85}.
A correlation between $A_V$ and the annular-averaged column
density of HI has been observed in both NGC~2403 \citep{mlm84}
and IC~342 \citep{mlm89}, substantiating the above claims.
Using a sample of giant
extragalactic HII regions in late-type spirals,
we construct the dust--gas relation
and use it to determine the extragalactic extinction of two
HII regions in Maffei~2.
By subtracting the extragalactic extinction from the total extinction
of the HII regions derived from
their Balmer decrements,
we obtain a measurement of the Galactic extinction.
In deriving the Galactic
extinction in various bandpasses,
we account for shifts in the effective wavelengths
of broadband filters, which can be significant for highly
reddened galaxies \citep[see][]{mlm04}.
To check the validity of our result,
we use the method applied to Maffei~2 to
obtain the Galactic extinction of the nearby
late-type spiral IC~342,
for which an accurate independent estimate is available
owing to its more moderate obscuration.
Armed with the extinction to Maffei~2,
we determine its distance using the Tully-Fisher
relation constructed in the $I$-band.

Our observations of HII regions in Maffei~2 and IC~342
are presented in
\S~\ref{observations} and reductions are outlined in
\S~\ref{reductions}.
In \S~\ref{measurements}, we describe the measurements
of the total extinctions
(Galactic plus extragalactic) and the hydrogen
column densities for the
observed HII regions.
In \S~\ref{extinction}, we construct the relation between dust and gas
and use it to determine the Galactic extinction of Maffei~2
and IC~342.
In \S~\ref{distance}, we re-determine zero-points for the extragalactic
distance scale in a self-consistent fashion.
We then derive the Tully-Fisher distance to Maffei~2
as well as a revised distance to IC~342 using recent observations
of Cepheids by \citet{sch02}.
The revised properties of both galaxies in addition to Maffei~1
are presented in
\S~\ref{implications} and the implications for the Local Group
are discussed.

\section{Observations}
\label{observations}

Long--slit spectroscopic observations of two HII regions in Maffei~2 
were obtained on 1997 March 2 and 3 UT with the
Ritchey--Chr\'{e}tien Spectrograph on the Mayall 4--m telescope at
Kitt Peak National Observatory.    
Observations were
optimized to detect both \ha\ and
\hb\ emission to determine the
extinction toward Maffei~2.
The T2KB 2048$\times$2048 CCD with 24~$\mu$m pixels was used in 
combination with the KPC10-A grating for wavelength coverage 
between 3500~\AA\ and 7600~\AA\ at 7~\AA\ resolution.
The spatial scale at the focal plane was 0\farcs{69} per pixel.
The slit was set to a width of 2\arcsec\ and aligned in the
north--south direction.
The slit was placed over the southwest portion of Maffei~2 at the
position of the brightest HII region \citep[][No. 1a]{spi73}.
Two individual clumps were also detected within a second HII region
at the position of object number 2 in \citet{spi73}.
Two 1800--second exposures were taken on the night of March 2,
and an additional three 1800--second exposures were obtained
the following night.
The effective air masses for the first and second night
were 1.65 and 1.54, respectively.

Inspection of a single 1800--second dark frame showed that the total dark
count rate was negligible compared to the sky.
Internal and twilight flats were obtained to correct for variations
over small and large spatial scales, respectively, on the CCD.
HeNeAr arc--lamp spectra were used for wavelength calibration.
Flux calibrations were achieved by observing the standard
stars Feige~67 and G191B2B \citep{oke90}.
Standard star exposures were interspersed among object exposures
over a range of airmasses spanning those of Maffei~2.

In addition to the Maffei 2 observations, spectrophotometry of
two HII regions in IC~342 was carried out to augment existing data for
that galaxy.
The spectra were acquired
on 1992 January 28 UT with the 2.3-m f/9 Bok telescope 
and Boller and Chivens Spectrograph
at Steward Observatory. The detector was a Texas Instruments 
$800 \times 800$ CCD with
$15 \, \rm \mu m$ square pixels. Observations
were made in first order
with a 300 lines~$\rm mm^{-1}$ grating
blazed at $6690 \, \rm \AA$. The grating was
rotated to give coverage from 4300 to $7200 \, \rm \AA$, over which
the dispersion was $3.7 \, \rm \AA / pix$.
The slit was opened up to $4\farcs 5$
to enable absolute calibration of line fluxes
and to negate the possibility of errors in line ratios
as a result of differential refraction.  Precise pointing was
guaranteed by rotating the slit to a position angle ($21^\circ$) which
admitted the light of a reference star only when the HII region
was centered. A single 1800-second exposure  was acquired, in which
$\rm H\alpha$ and $\rm [N~II]\lambda 6584$ were clearly resolved.
In order to minimize the effects of flexure on flat fielding,
a spectrum of an internal quartz lamp was acquired at the same location
immediately following the target exposure. 
A helium-argon lamp was observed at a comparable position to
calibrate the wavelength scale.
A spectrum of twilight was acquired at the beginning of the night
to map out the illumination pattern.  To calibrate fluxes,
the standard stars Hiltner 102, Feige 15, G191B2B, and HD~84937
were observed with the same wide slit as employed for the HII regions.

\section{Reductions}
\label{reductions}

The Maffei~2 spectra were reduced using standard IRAF\footnote{IRAF is
distributed by the National Optical Astronomy Observatories, which is
operated by the Associated Universities for Research in Astronomy,
Inc., under contract to the National Science Foundation.}
routines.
Reductions were carried out separately for each night.
The object and flat field frames were overscan--corrected and
bias--corrected.
The internal flat and sky flat frames were combined to produce a
single internal flat frame and a single averaged sky flat frame.
The averaged sky flat frame was corrected for pixel-to-pixel
variations in response using a normalized internal
flat frame. 
An illumination image was created from the processed sky flat to
correct for the slit function. 
Then, each object frame was corrected by dividing the product
of the illumination image and normalized internal flat.
With two-dimensional fits to the arc spectra,
geometric distortions were corrected so that the dispersion axis was
made perpendicular to the spatial axis.
Spectra of the two standard stars were combined to
calibrate fluxes.
Air mass corrections were derived using a mean
atmospheric extinction curve for KPNO.
The five Maffei~2 spectral exposures were then aligned, shifted, and
combined into a single spectral frame.
The IC~342 spectra were reduced using standard IRAF routines
in a similar manner as for the Maffei~2 spectra.
Final one-dimensional flux-calibrated
spectra were obtained via unweighted summed extractions.

Table~\ref{tablines} lists the observed flux ratios
relative to $\hb$ for the HII regions observed in Maffei~2 and IC~342.

\section{Measurements for Maffei~2 and IC~342}
\label{measurements}

\subsection{Total Extinction}
\label{s_extot}

The extinction of a source is best quantified by its optical depth
at some wavelength.
Knowing optical depth, it is possible to evaluate the extinction
in any filter
by applying an appropriately scaled reddening law to the spectrum
and integrating through
the response function for the filter.
The best choice of wavelength is 1~$\mu$m, as it falls in a part
of the reddening law which is not very sensitive to environment,
and because the optical
depth there is comparable numerically to $E(B-V)$ \citep[see][]{mlm04}.

The total extinction (Galactic plus extragalactic)
of an HII region can be determined from
the degree to which the observed $\hab$
ratio differs
from the theoretical intrinsic ratio.
The optical depth at 1~$\mu$m ($\tau_1$)
can be computed from
\begin{equation}
\label{eqtau}
\tau_1 = 2.5 [\log{(\fab)^0}-\log{(\fab)}]/(R^1_{\ha} - R^1_{\hb})
\end{equation}
where $(\fab)^0$ is the intrinsic Balmer ratio,
\fab\ is the observed Balmer ratio corrected for underlying
stellar absorption,
and $R^1_{\ha}$ and $R^1_{\hb}$ are the reddening coefficients
normalized to $\tau_1$ (i.e., $R^1_\lambda=A_\lambda/\tau_1$,
where $A_\lambda$ is the extinction corresponding to $\tau_1$
at the wavelength $\lambda$).
To estimate the reddening coefficients,
we require a monochromatic extinction curve appropriate for
the diffuse interstellar medium, which is primarily responsible
for obscuring extragalactic sources.
\citet{fi99} has developed an algorithm for determining
the monochromatic reddening law associated with any particular
value of the ratio of total to selective extinction, $R_V=A_V/E(B-V)$.
For the diffuse interstellar medium, $R_V = 3.07 \pm 0.05$
for a star of zero colour in the limit of zero extinction \citep{mlm04}.
The reddening law chosen to evaluate the reddening coefficients
is that given by the
algorithm of \citet{fi99} which, when applied to the spectrum of Vega,
yields $R_V = 3.07$
after integrating the flux passed by response curves
characterizing the $B$ and $V$ filters.
The resulting Balmer coefficients are $R^1_{\ha}=2.185$ and $R^1_{\hb}=3.374$.

The intrinsic Balmer ratio $(\fab)^0$ for each HII region
was estimated from the emissivities of
\citet{sh95}.
An initial approximation for \tauone\ was derived from Eq.~\ref{eqtau}
using $(\fab)^0=2.91$, which was computed
by assuming a
temperature of 8000~K and an electron density of 100~$\rm cm^{-3}$,
these being typical values for giant extragalactic HII regions
\citep{mrs85}.
Then, the approximation for \tauone\ was used
to correct \oii/\hb\
and \oiiib/\hb\ for reddening
using reddening coefficients
$R^1_{\rm 3727}=4.379$ and $R^1_{\rm 5007}=3.237$ computed
in the same manner as for \ha\ and \hb.
The corrected oxygen line ratios were used to estimate
the oxygen abundance, $\rm \ox=\log{n(O)/n(H)}$,
via the semi-empirical calibration
of \citet{mcg97}.
Next, \ox\ was used to obtain an improved approximation of the
HII region temperature using the
theoretical correlations between electron temperature
and oxygen abundance formulated by \citet{mcg91}.
The correlations are parameterized by the
volume-averaged ionization parameter ($U$) and
the upper mass limit for the ionizing stars.
However, in the oxygen abundance range inhabited by
the reference HII regions ($-3.9 < \ox < -2.8$),
the relations converge to straight lines of nearly-identical
slope with a spread of approximately $\pm600$~K for intermediate
values of $U$, which are typical of observed HII regions
\citep{mcg91}. Thus, the temperature (T) in Kelvins is given by
\begin{equation}
\label{eqtemp}
\rm T = -7390 \ox - 17433
\end{equation}
Eq.~\ref{eqtemp}
was used to derive the next approximation of $(\fab)^0$,
which was then substituted back into
Eq.~\ref{eqtau} to obtain the next estimate of \tauone.
This process was iterated until $(\fab)^0$,
and thereby \tauone, converged.
Table~\ref{tabhiiex} lists the observed and intrinsic fluxes
as well as the oxygen abundances, temperatures and total extinctions
found for the HII regions in Maffei~2 and IC~342.

\subsection{Hydrogen Column Densities}
\label{s_gas}

The surface brightnesses
of HI 21-cm and CO 2.6-mm line radiation are traced
by atomic and molecular hydrogen gas, respectively,
which is distributed all the way
from the far side to the
near side of a galaxy. In the Milky Way, the scale heights
of the HI and $\hmol$ gas are $\sim150$~pc and $\sim60$~pc,
respectively \citep{mal94,mal95}.
The HII regions, being
associated with the young stellar population, are confined
to a layer with a scale height of only $\sim90$~pc \citep{kru96}.
The column density of
extragalactic gas in the foreground of an HII region can therefore be
approximated as half the column density of hydrogen derived
from observations of HI and CO along the line of sight, with some
scatter caused by deviations in position from the
exact mid-plane and by concentrations of dust in the immediate vicinity
of the HII region.
Thus, the hydrogen column density $\nh$ in atoms $\rm cm^{-2}$
along the line of sight is given by
\begin{equation}
\label{eqnh}
\nh = N_{\rm H}/2 = (\nhi+2\nhmol)/2
\end{equation}
where \nhi\ and \nhmol\ are the column densities of HI and \hmol\,
respectively.
The \nhmol\ term
is doubled to preserve the proportionality between the
number of dust particles and hydrogen atoms.

The annular-averaged column densities of HI 
at the de-projected radii of the two HII regions in Maffei~2
were measured from the HI radial profile
of \citet{hth96}.
The CO (J=1-0) intensities at the locations of the HII regions
were obtained from the radial profile provided
by Mason \& Wilson (private communication; see also \citealt{mw04}).
For IC~342, the HI column densities and CO intensities at the locations
of the HII regions were measured from the
radial profiles of \citet{new80b} and \citet{cth01}, respectively.

The conversion from CO intensity ($\ico$) to $\hmol$ column density
was found to have a metallicity dependence
in independent studies by \citet{wil95} and \citet{ast96}.
The conversion relation adopted here is from the
former source, in which measurements of $X = \nhmol/\ico$ and
$\ox$ are determined in a homogeneous manner by applying
the virial theorem to individual molecular clouds in nearby spiral
and dwarf irregular galaxies.
For each HII region,
the conversion factor ($X$) along with the
HI and $\hmol$ column densities are summarized in Table~\ref{tabhiiex}.

\section{Extinction}
\label{extinction}

\subsection{The Relationship Between Dust and Gas}
\label{s_gasdust}

We have constructed the relationship between extragalactic extinction and
hydrogen column density for a sample of 74 reference HII regions
in 10 spirals of type
Sbc and Scd.
We have restricted our sample to galaxies for which
(1) the Galactic extinction
is small and thus well-known (i.e., $\ebv < 0.05$~mag);
(2) the radial profile of HI has been determined from
aperture synthesis maps;
and (3) there is a published radial profile of the CO (J=1-0) line.
We have further required that the constituent HII regions be
non-nuclear to insure the reliability of abundance calibrations,
where ``non-nuclear'' is defined here as having a radius
greater than 15\% of the host galaxy's disk scale length.
In addition, we have rejected any HII region located beyond the
radial boundaries of the HI maps. 
Lastly, all HII regions were required to have measurements
of the flux at \ha\ and \hb\
as well as
the equivalent width of the \hb\ emission line
in order to account for the depression of Balmer
emission by underlying Balmer absorption from stars
(for typical HII regions, the equivalent width
of \ha\ emission is so strong that the correction to the \ha\ flux
is negligible).
The galaxies meeting the above criteria are listed
in Table~\ref{tabspi}.

The total extinctions of the reference HII regions
were derived in a manner consistent with those
employed for Maffei~2 and IC~342 (see \S~\ref{s_extot}).
Several of the HII regions were missing measurements for the
oxygen line fluxes as a consequence of poor signal-to-noise.
Their temperatures were determined from a correlation between
temperature and galactocentric radius found for
the HII regions for which oxygen line fluxes were available.
As can be seen in Fig.~\ref{figtr},
the slope and intercept of the correlation
appear to have a dependence on spiral morphology 
in the sense that HII regions in late-type spirals
have higher temperatures.
We therefore adopt two linear fits; one for the
spirals of type Sbc and one for those of type Scd.
The fits are described by
\begin{equation}
\label{eqtsbc}
T_{\rm Sbc} = (1109 \pm 137) r/r_0 + (3004 \pm 266)
\end{equation}
\begin{equation}
\label{eqtscd}
T_{\rm Scd} = (656 \pm 171) r/r_0 + (6026 \pm 367)
\end{equation}
where T is in Kelvins and $r/r_0$ is the deprojected galactocentric distance
of the HII region normalized to the host galaxy's disk scale length.
The root-mean-square (rms) deviations for the Sbc and Scd samples
are 562~K and 934~K, respectively.

The extragalactic extinction of each HII region (quantified by \tauex)
was extracted from
the total extinction (\tautot) observed in the Balmer decrement
(Eq.~\ref{eqtau})
by subtracting the galactic extinction (\taugal) derived
from the Galactic reddenings of \citet[][hereafter SFD]{sfd98}.
For each host galaxy,
we computed \taugal\ from $E(B-V)$ using the iterative process outlined in
\citet{mlm04}. Briefly,
the integrated spectral energy distribution (SED) of a typical unreddened elliptical galaxy
was extinguished by successive approximations of \taugal\ with the aid
of the scaled monochromatic reddening law described in \S~\ref{s_extot}.
The application of this process was greatly facilitated by
the York Extinction Solver~(YES).\footnote{
The York Extinction Solver~(YES) is a web-based application
developed by the Department of Physics and Astronomy, York University
and hosted by the Canadian Astronomy Data Centre (CADC).
It can be accessed at http://cadcwww.hia.nrc.ca/yes.}

The total gas column density ($\nh$) at the location of each
reference HII region was determined
according to the procedure outlined in \S~\ref{s_gas}.
The gas column densities and extinctions of the HII regions
are provided in Table~\ref{tabhii} along with the sources of
the HI and CO data.
The correlation between \tauex\ and $\nh$
is shown in Fig.~\ref{figdust}
and does not appear to depend
on spiral morphology within the range Sbc to Scd.
We have therefore adopted the following linear least-squares fit to
the entire reference dataset:
\begin{equation}
\label{eqdust}
\tauex = (\mtau) \nh + (\btau)
\end{equation}
where \nh\ is in units of $10^{20} \rm \, cm^{-2}$.
The rms deviation in \tauex\ is \rmstau.

Eq.~\ref{eqdust} improves upon the dust--gas relation found previously
by \citet{mlm89} by expanding the HII region sample to include
recent observations, by excluding NGC~6946 due to its high
Galactic extinction, and by including $\hmol$
in the gas diagnostic. This last step has reduced the scatter
considerably.

\subsection{Uncertainties in the Dust--Gas Relation}
\label{s_err}

The error in an estimate of \tauex\ for an HII region
from Eq.~\ref{eqdust}
can be computed from
\begin{equation}
\label{eqdusterr}
(\delta \tauex)^2 = (m \, \delta \nh)^2 + [\delta m (\nh - \langle\nh\rangle)]^2
+ \sigma_{\tauone}^2/n
\end{equation}
$\delta m$ is the standard error associated with $m$,
the slope of the dust--gas correlation (Eq.~\ref{eqdust}).
$\delta  \nh $ is the measurement error associated with the estimate
of $\nh$ for the HII region.
$\sigma_{\tauone}$ is the rms deviation of \rmstau\ in \tauex\ of the
$n=74$ reference data points about the linear fit.
$\langle\nh\rangle$ is the mean \nh\ for the HII region sample,
found to be $\avenh \times 10^{20} \rm \, cm^{-2}$.

The measurement uncertainty in \nh\ for each reference HII region was computed
from one half of the quadrature sum of the errors in \nhi\ and 2\nhmol.
The greatest source of uncertainty in a measurement of \nh\
arises from taking the annular average of intensity measurements
at a given radius. The uncertainty increases with distance from
the nucleus; the larger the circumference of an annulus, the larger
the area of galaxy contained within the annulus, and therefore the
greater the possibility of intensity fluctuations.
We have derived an
expression for this uncertainty by examining the
reference galaxies for which
measurements of HI were supplied separately for each half of the disk.
For a given annulus, the
difference between the column densities within each semi-annulus
($\Delta \nhi$) relative to the mean ($\langle\nhi\rangle$)
is found to be a linear function of the fractional galactocentric
radius $r/r_0$, with an rms deviation of \rmsdnh.
The uncertainty in a measurement of \nhi\ from
an annular-averaged radial distribution can therefore
be estimated from
\begin{equation}
\label{eqnerr}
\Delta \nhi / \langle\nhi\rangle = 0.04 r/r_0 + 0.08
\end{equation}
Adopting this approach,
the average error in \nhi\ for the reference HII regions is \avednhi\%.

The dominant contributions to the uncertainty in \nhmol\
for each reference HII region
are the measurement uncertainty in \ico,
for which we find a mean value of \avedico\%,
and the uncertainty in the CO--to--\hmol
conversion factor ($X$).
This latter source is dominated
by the scatter of $1.03 \times 10^{20} \rm \, cm^{-2} (K \, \kms)^{-1}$
in the adopted metallicity relation (see \S~\ref{s_gas})
and the measurement uncertainty in the
oxygen abundance, for which we adopt 0.2 dex as recommended
by \citet{mcg91}.
Taking the above sources into account,
we find a mean measurement uncertainty in \nh\ of 19\%
for the reference HII regions.
We note that the mean measurement uncertainty in \nh\
for the Sbc galaxies exceeds that found for the Scd galaxies
by 6\%, as a result of the
considerably larger uncertainties in \nhmol.
This is likely due to the higher \hmol\ content
in the Sbc galaxies,
as nearly 40\% of the Scd HII regions are found in regions of
negligible CO.

The measurement uncertainty in \tauex\ for each HII region is simply
the quadrature sum of the uncertainties associated with \tautot\ and \taugal.
The uncertainty in \tautot\
is dominated by the measurement
error in the observed \hab\ flux,
given the weak dependence of the intrinsic Balmer ratio on temperature.
(The mean error in $(\fab)^0$ owing to the
temperature estimate was found to be under 2\%
and was therefore considered negligible.)
The mean measurement error in \fab\ is \avedfab\%,
which was computed from
the quadrature sum of the signal-to-noise error in the flux
and a 10\% error arising from the uncertainty in the
overall response correction \citep[][p.103]{mlm82}.
The contributions from the uncertainties in the
reddening coefficients were found to be negligible.
Adopting a 0.16 fractional error in \taugal\
as recommended by SFD,
the mean measurement uncertainty in
\tauex\ for the reference HII regions is \avedtauex.
Thus, the scatter of \rmstau\ in the dust--gas relation
may be mostly due to the measurement uncertainties.
The scatter above that associated with the measurement errors
may be arising from dust concentrated near the HII regions,
which is not traced by $\nh$.
The non-zero intercept proves the existence of such localized dust.
However, the fact that this scatter
does not swamp the
dust--gas trend implies that the extinction due to this dust
is roughly constant among HII regions.
A value of \tauex\ obtained from Eq.~\ref{eqdust}
therefore includes both the optical depth due to widespread dust in
the extragalactic foreground as well as the average optical depth
due to dust associated with the star formation regions.

\subsection{The Galactic Extinction of Maffei~2 and IC~342}
\label{s_ext}

The derived extragalactic extinctions
for the HII regions in Maffei~2 and
IC~342 are given in Table~\ref{tabhiiex}.
The quoted errors were derived in the same manner
as for the reference HII regions (see \S~\ref{s_err}).
We find mean total extinctions of $\tautot=\tautotmafii \pm \dtautotmafii$
for Maffei~2
and $\tautot = \tautotic \pm \dtautotic$ for IC~342.
\citet{mf92} estimated the total reddening of
hot stars in IC~342
from a comparison of the $B-V$ colour of the blue ``plume''
in the colour-magnitude diagram
with that of IC~1613.
They obtained $E(B-V) = 0.79 \pm 0.05$~mag,
corresponding to $\tautot = 0.82 \pm 0.05$
based on the SED of a B0 V star.
Their result is consistent within errors with the mean total
extinction found for HII regions in IC~342.

In Fig.~\ref{figdust}, the total extinctions for the heavily obscured
HII regions
are plotted over the dust--gas correlation found for the reference
HII regions.
The weighted mean of the extinction offsets of the heavily
obscured HII regions imply Galactic extinctions of
$\taugal = \taumafii \pm \dtaumafii$ for Maffei~2 and
$\taugal = \taucic \pm \dtaucic$ for IC~342.

The SFD reddening for IC~342 is $E(B-V)=0.558 \pm 0.069$~mag,
which corresponds to a Galactic optical depth of
$\taugal = \tausfdic \pm \dtausfdic$ according to the treatment
outlined in \S~\ref{s_gasdust}.
This value is consistent with our estimate of
$\taucic \pm \dtaucic$ within errors,
which validates our extinction analysis in general as well
as the extinction estimate for Maffei~2 in particular.
While SFD caution against the use of their reddening
maps for objects within $\pm5^\circ$ of the Galactic
plane, the Galactic latitude of IC~342 is safely outside this range
at $b=10^\circ.58$.
We therefore have no reason to reject the SFD value of \taugal\ for this
galaxy, so we adopt the weighted mean of both values,
which yields $\taugal=\tauic \pm \dtauic$.

The extinction parameters and corrected magnitudes of Maffei~2 and IC~342
are given in Table~\ref{tabprop}.
The lower values of the reddening coefficients $A_\Lambda/\tauone$
observed for Maffei~2 for each broadband filter $\Lambda$
are due to shifts in the effective wavelengths of $\Lambda$
towards the red caused by
the higher Galactic extinction and inclination.

\section{Distances}
\label{distance}

\subsection{Zero-points of the Extragalactic Distance Scale}
\label{s_zp}

As explained in \S~\ref{s_extot} and \S~\ref{s_gasdust},
analyses in this paper are founded upon a modern
framework for handling extinction which
eliminates the biases
that accompany traditional approaches.
As a consequence, zero-points for
distance indicators employed in the HST Key Project
on the extragalactic distance scale
(hereafter \citealt{fre01})
must be updated using the same
approach.  Specifically, revised
distances are required to the Ursa Major cluster, which defines the
Tully-Fisher (TF) relation employed here, and to the Coma cluster,
which defines the Fundamental Plane (FP) used to determine the
distance of Maffei 1 in \citet{fmd03}.

Behind the TF and FP calibrations
are Cepheid distances to nearby galaxies, which in the
\citet{fre01} are anchored to the Large Magellanic Cloud
(LMC).  As a result,
the period-luminosity (PL)
relations for the LMC must be re-visited.
The \citet{fre01} PL relations come from
\citet{ud99a}, and the extinction
corrections inherent to the PL relations,
which are based on the apparent magnitude of
the red clump,
are those
of \citet{ud99b}.
\citet{ud99b} presumed that the red clump has a constant
absolute magnitude in Cousins $I$ (henceforth $I_C$), 
and that changes in apparent magnitude are due
to fluctuations in extinction with respect
to some zero-point.
The zero-point was determined from colour-magnitude diagrams
for an eclipsing binary (HV 2274) and 
for two young open clusters.
It appears that a reddening law for the LMC 
\citep{fi85} was used to define
$E(B-V)$ for each of these sources \citep[see][]{ud98}, but
the determination of corrections to $E(B-V)$ from 
red-clump magnitude fluctuations
and the conversion of total reddenings to extinctions in $V$ and $I_C$
were based upon extinction coefficients tabulated by
\citet{sfd98}, which are founded upon the broadband
reddening law of \citet{ccm89} for the Milky Way.

Fortunately,
the mean reddening of the Cepheids behind the \citet{fre01} PL
relations, $E(B-V) = 0.147$~mag (range 0.11 to 0.20 mag), is
very close to the value measured for the objects
defining the zero-point (0.13 to 0.15 mag). This means that red-clump
adjustments to the Cepheid extinctions
amount to second-order corrections,
and any errors in these adjustments
ought to average out.
Thus, red clump stars are of less concern than the
matter of what sets the zero-point.
The reddening that \citet{ud99b} adopted for the binary
was $0.149 \pm 0.015$~mag \citep{ud98}.  However, 
\citet{gui98}
derived $E(B-V) = 0.120 \pm 0.009$~mag by simultaneously solving
for atmospheric temperatures and the interstellar extinction curve.
In making their determination, \citet{gui98} incorporated 
photometry from \citet{ud98}. This removed the 
reddening degeneracy
which afflicted a preliminary determination criticized by 
\citet{ud98}.  The result of \citet{gui98}
has to be taken very seriously because 
the SED, reddening law, and colour excess are not only
consistent with the LMC environment, but also consistent with each
other. It is concluded that
the $E(B-V)$ scale of \citet{ud99b}
is too high by $0.029 \, \rm mag$.

Correcting for the change in the reddening zero-point, one gets
$E(B-V) = 0.118$~mag
for the mean reddening of LMC Cepheid fields as tracked
by early-type stars.  This is close to the
value (0.10 mag) adopted by \citet{mf91}.

According to \citet{sfd98}, $E(B-V) = 0.075$~mag
is the Galactic reddening based upon what is
observed in annuli around the LMC.  With the rest-frame
elliptical SED of \citet{mlm04}
and the reddening law of \citet{fi99}
tuned to give $A_V / E(B-V) = 3.07$ for Vega,
the optical depth of Galactic dust toward
the LMC at $1 \, \rm \mu m$ is $\taugal = 0.085$.  
Adopting the SED
B12 III of \citet{pic98} for the stars in HV~2274, 
a heliocentric velocity of 265~\kms
\citep{fio83} and the reddening law of
\citet{gor03} for dust inside the LMC, 
the mean value $E(B-V) = 0.118$~mag for the
Cepheid fields leads to an optical depth $\tauex = 0.040$ for
dust ``localized'' in the LMC.
In turn, applying the redshift and optical depths to SEDs
of Cepheids, which here are adopted to be G0 supergiants based upon the
periods (there is hardly any difference in results between F8 I
and G2 I), one gets the following sums of Galactic extinction,
localized extinction, and the difference in K-corrections in $V$ and $I_C$:
$A_V = 0.361$~mag (versus 0.476 mag from \citealt{ud99b}),
$A_{I_C} = 0.202$~mag (versus 0.288 mag from \citealt{ud99b}),
and $E(V-I_C) = 0.159$~mag (versus 0.147 mag from \citealt{ud99b}).
This means that the zero-points for the PL relations for the 
LMC Cepheids change as follows.  For a distance modulus of
18.50~mag, 
\begin{eqnarray}
M_V &=& -2.760 (\log P - 1) - 4.103 \\
M_{I_C} &=& -2.962 (\log P - 1) - 4.818 \\
V - I_C &=& 0.202 (\log P - 1) + 0.715
\end{eqnarray}
where P is the period in days.

As the starting point for correcting
\citet{fre01} Cepheid distances,
apparent distance moduli $\mu_V$ in $V$ and $\mu_{I_C}$ in 
$I_C$ recorded in the \citet{fre01}
were adopted.
First, all were corrected to account for the new
PL relations for the LMC (i.e., the revised extinction zero-point
for LMC Cepheids).  Next, foreground
values of $E(B-V)$ from \citet{sfd98} were converted
to \taugal\ by applying the reddening law of \citet{fi99}
(tuned to give $R_V=3.07$ for Vega) 
to the rest-frame elliptical SED of \citet{mlm04}.  Values of
\tauex\ were derived individually for each galaxy from 
$\mu_V -\mu_{I_C}$, the redshift, and the SED of a G0 supergiant (a Cepheid),
again using the reddening law of \citet{fi99}.
Then, values of Galactic extinction, $A_{I_C}^{\rm gal}$, localized
extinction, $A_{I_C}^{\rm loc}$, and the K-correction,
$K_{I_C}$, were computed.  The corrections were
subtracted from $\mu_{I_C}$ to arrive at corrected distance moduli
$\mu^0$ on a scale where the distance modulus of the LMC is 18.50 mag.
Note that $K_V$ can reach as high as 0.009 mag for a Virgo Cluster galaxy,
so K-corrections
are just starting to become relevant [$K_V$, of course, affects
the determination of \tauex\ via $E(V-I_C)$].
Despite all the revisions, Cepheid distances increased by
only $0.007 \pm .005$~mag on average.

As in the \citet{fre01},
metallicity-corrected distance moduli $\mu^0_Z$ were computed
from oxygen abundances $\rm \log Z = 12 + \log{n(O)/n(H)}$
estimated from HII regions via the calibration of
$R_{23}=(\oii+\oiiia+\oiiib)/\hb$
provided by \citet{zar94}.  Formally,
\begin{equation}
\mu^0_Z = \mu^0 + \delta_{VI}
\end{equation}
\noindent
where
\begin{equation}
\delta_{VI} = \gamma_{VI} ( \log Z_{LMC} - \log Z)
\end{equation}
\noindent
From \citet{sak04},
$\gamma_{VI} = -0.24$ and
$\log Z_{LMC} = 8.50$.
The adopted value of $\gamma_{VI}$ is slightly more negative than that
employed in the \citet{fre01}.

\citet{sak00} specify the Cepheid
calibrators used to establish the zero-point for the
TF relation and in turn distances to the Ursa
Major and Coma clusters.  Besides correcting the Cepheid distances,
magnitudes for both the calibrators and the cluster
galaxies 
must be re-computed using a self-consistent
approach to extinction.  To make this possible,
S. Sakai (personal communication) provided
the apparent magnitudes and corrections employed in
the \citet{fre01}.
In the process, an error in the \citet{fre01}
Galactic extinction corrections was found; the $I_C$-band
correction for all TF galaxies was accidentally set
to 59\% of the desired value.

For each Cepheid calibrator, new estimates of the 
Galactic extinction and the K-correction
were computed from the value of $E(B-V)$ given by \citet{sfd98}
and the redshift using the SED of \citet{mlm04} closest to the
Revised Hubble type of the calibrator, adjusted appropriately
for tilt.
The internal extinction corrections of \citet{sak00} were retained.
With weights set by the random uncertainties in
the absolute magnitudes, the mean offset between the new and old 
absolute magnitudes amounts to
\begin{equation}
M_{I_C} - M_{I_C} (\hbox{\citealt{sak00}}) = +0.048 \rm \, mag
\end{equation}

Improved Galactic extinction and K-corrections for 
Ursa Major galaxies were computed from the mean value of 
$E(B-V)$ (from \citealt{sfd98}) and redshift, 
which were judged from the mean cluster coordinates and velocity
recorded by \citet{sak00}.  To this end, the Sbc SED
of \citet{mlm04} was adopted, adjusted to the mean tilt and luminosity
of the local calibrators.
For consistency with the calibrators,
the corrections for internal extinction
adopted by \citet{sak00} were retained.
With $E(B-V) = 0.025$~mag and
a heliocentric velocity of 899 \kms, 
the weighted mean offset between the new and old 
apparent magnitudes corrected for Galactic extinction,
redshift, and internal extinction amounts to
\begin{equation}
m^0_{I_C} (\hbox{UMa}) - m^0_{I_C} (\hbox{\citealt{sak00}}) = -0.017 \rm \, mag
\end{equation}
\noindent
The revisions to apparent and absolute magnitudes
lead to an offset in the distance modulus for Ursa Major of
\begin{equation}
\mu^0_Z (\hbox{UMa}) - \mu^0 (\hbox{\citealt{sak00}}) = -0.065 \rm \, mag
\end{equation}

The distance modulus for Ursa Major 
determined by \citet{sak00} was
31.58 mag, so the revised distance modulus comes out to be
$31.515 \pm 0.13$~mag for an LMC distance modulus of 18.50 mag.  
The quoted error is the random error in the zero-point of the
TF relation.

The distance to the Coma cluster adopted in this paper is the mean of
results from the TF relation and the FP.
The revision to the TF distance for Coma was undertaken
in exactly the same manner as that for Ursa Major. Apparent magnitudes
shift as follows:
\begin{equation}
m^0_{I_C} (\hbox{Coma}) - m^0_{I_C} (\hbox{\citealt{sak00}}) = -0.015 \rm \, mag
\end{equation}
\noindent
Combining this with the absolute magnitude shift computed above
for the Cepheid calibrators, the distance modulus shifts by
\begin{equation}
\mu^0_Z (\hbox{Coma}) - \mu^0 (\hbox{\citealt{sak00}}) = -0.063 \rm \, mag
\end{equation}
\noindent
Since \citet{sak00} derived a distance modulus of 34.74 mag,
the new distance modulus for Coma works out to be $34.677 \pm 0.13$~mag, where
the uncertainty is the random error in the zero-point of the
TF relation.

The FP analysis for the \citet{fre01} was
conducted by \citet{kel00}.  For any particular galaxy,
the zero-point of the FP is defined by
\begin{equation}
\gamma = \log r_e - 1.24 \log \sigma + 0.82 \log < I >_e
\end{equation}
\noindent
where $r_e$ is the effective metric radius (of an $r^{1/4}$ law),
$\sigma$ is the velocity dispersion, and $< I >_e$ is the
mean surface brightness within the effective radius.
Of course, one does not measure $r_e$, but rather $\theta_e$,
the effective radius in angular units.  If $r_e$ is replaced
by $\theta_e$, variations
in $\gamma$ from galaxy to galaxy 
arise from variations in distance.  The fiducial
value of $\gamma$ was defined using the
Leo Group and the
Virgo and Fornax clusters, in which Cepheid 
calibrators are located.  
Revisions to extinction or to K-corrections affect
$\gamma$ through $< I >_e$ and through $r_e$, the latter 
being determined by the distances assigned to the reference clusters
(which depend upon the treatment of Cepheids).

Photometry in $V$ was used to define $\gamma$
for Leo, Virgo and Fornax, and photometry in Gunn $r$ was used
to define $\gamma$ for the Coma cluster.
Revising extinction estimates as recommended by \citet{mlm04},
and computing K-corrections self-consistently, the changes to
$< I >_e$ decrease the weighted mean value of $\gamma$ for
the calibrating clusters by $0.0013$ and decrease the value
of $\gamma$ for Coma by $0.0065$.  The revised Cepheid analysis
described above moves the distance moduli for Leo, Virgo and
Fornax farther away by $0.012 \, \rm mag$ on average with respect
to the distance moduli adopted in the \citet{fre01}.
In this computation, clusters were weighted on the basis of
the uncertainty in $\gamma$ \citep{kel00}.
The combination of all effects leads to an increase 
in the distance modulus of the Coma cluster by $0.038 \, \rm mag$
with respect to the \citet{fre01}.  On a scale where the
distance modulus of the LMC is 18.50 mag, the revised distance modulus to
Coma becomes $34.705 \pm 0.15$~mag, where the uncertainty is that
due to random errors.  

The distance moduli for Coma derived from the TF
relation and the FP differ by only 0.028 mag.
We adopt the unweighted average,
which is $34.69 \pm 0.10$~mag.

In this paper, the authors prefer to adopt the maser distance
to NGC 4258 as the zero-point for distances, rather than the
LMC.  This galaxy was among the Cepheid calibrators in the \citet{fre01}.
From the re-analysis above,
$\mu_Z^0 (\hbox{N4258}) = 29.526$~mag
with a random error of 0.07 mag.  
This is 0.016 mag higher than the \citet{fre01} value.
From \citet[][see also \citealt{gib00}]{hmg99},
$\mu (\hbox{masers}) = 29.29$~mag, so
\begin{equation}
\mu(\hbox{masers}) - \mu(\hbox{Cepheids}) = -0.236 \rm \, mag
\end{equation}

\noindent
Referencing to the maser zero-point, 
the distance moduli for Ursa Major
and Coma are $\muuma \pm \dmuuma$~mag
and $\mucoma \pm \dmucoma$~mag, respectively.

A proper treatment of the shift in the \citet{fre01} Hubble Constant (\ho)
due to the above extinction corrections would necessitate a re-analysis
of the zero-points for $all$ distance indicators behind the measurement.
Since our distance measurements to Maffei~2 and IC~342 do not
depend on \ho, we do not require a value of \ho\ on the
extinction and distance scale adopted in this paper.
However, for the sake of the discussion in \S~\ref{implications}
in which we compare the new distance to the IC~342/Maffei~group
with that implied by the Hubble Flow, we adopt the \citet{fre01} value of
$\ho = 72 \, \kms$
after (1) correcting for the mean increase of $0.007 \pm 0.005$~mag
in the distance moduli of the Cepheid
calibrators due to the
extinction corrections described above;
and (2) correcting for the decrease of \delzp~mag in all \citet{fre01}
distance moduli due to the shift from the LMC
to the maser zero-point.
After applying these corrections, we obtain $\ho=\hub$.
This value is consistent with the most recent measurement of
$\rm 77\pm11 \,\kms\,Mpc^{-1}$ obtained
from the Sunyaev-Zel'dovich Effect measured by the Chandra X-Ray
Observatory for
high-redshift galaxy clusters \citep{bon06}, a result which
is completely independent of the \citet{fre01}.
In addition, \citet{cia02} use the planetary nebula luminosity function
(PNLF) to derive a distance to NGC~4258 which is in close agreement with
the maser distance and which increases the \citet{fre01} value of \ho\
to $\rm 78\pm7 \,\kms \, Mpc^{-1}$.
The agreement between these two studies substantiates
the distance scale adopted in this paper.

\subsection{The Distance to Maffei~2}
\label{s_dist_maf}

For a very long time,
studies of Maffei~2 were encumbered by the lack of an
apparent magnitude.
A reliable measurement of the total apparent magnitude
of Maffei~2 was finally made by \citet{bm99} in $I$,
which allows application of the TF relation to determine
the distance.
We require a TF relation in which the luminosity diagnostic
is derived from HI rotation curves as opposed to line widths,
owing to the contamination of Galactic HI in the HI line profile
of Maffei~2.
As a TF relation of this form was not available in $I$,
we constructed the relation
from the $I$-band photometry and HI synthesis observations
of spirals in the Ursa Major (UMa) cluster
conducted by \citet[][hereafter V01]{ver01}.
For the distance modulus of UMa, we adopted $\muuma \pm \dmuuma$~mag
(see \S~\ref{s_zp}).
Of the rotational velocity parameters measured by V01,
we chose the plateau rotational velocity (\vflat),
defined as the average amplitude of the flat outer region
of the rotation curve.
V01 found the least scatter in the TF relation constructed
from this diagnostic,
as opposed to the maximum observed rotational velocity.
We restricted our TF analysis to spirals
for which \vflat\ could be measured with confidence
(referred to by V01 as the RC/DF sample).
For consistency with the galaxy sample from which our adopted
UMa distance was derived,
we excluded the 6 galaxies from the RC/DF sample with I-band
tilt corrections greater than 0.6 mag.
We also excluded NGC~3992 based on the arguments of V01
that the galaxy may be in the background of UMa.
Our final sample of TF calibrators contains 15 spirals.

Great care was taken to correct the TF ingredients
in the same manner as for Maffei~2.
Galactic extinctions were computed from SFD reddenings
as described in \S~\ref{s_gasdust}.
The total $I$ magnitude of each galaxy was corrected to its
face-on value by employing YES to
solve for the internal extinction from the rotational velocity and the
apparent axis ratio
\citep[see][]{mlm04}.
K-corrections to the UMa photometry were found to be
negligible in $I$ (less than 0.01 mag).
The TF data are given in Table~\ref{tabtf}
and plotted in Fig.~\ref{figtf}.

To construct the TF relation,
we followed the treatment of \citet{fre01}
and determined the slope and zero point
using a bivariate linear fit,
minimizing errors in both $\log{2 \vflat}$ and $M_I$.
As in V01, we assumed that all galaxies have equal relative uncertainties
of 5\% in \vflat\ and equal photometric uncertainties of 0.05 mag in $M_I$.
The fit yields
\begin{equation}
\label{eqtf}
M_I=(\mtfi) \log{2 \vflat} + (\btfi)
\end{equation}
with an rms scatter of \rmstfi~mag in $M_I$.
The scatter is due in large part
to the depth of the UMa cluster;
the virial radius of 880 kpc \citep{tul96}
corresponds to a range of \spreaduma~mag in $M_I$,
which is comparable to the rms deviation over and above the measurement
uncertainties.

Our fit is in close agreement with V01,
who found a slope, zero-point and rms scatter
of $-10.4 \pm 0.4$, $4.27 \pm 0.89$~mag, and 0.30 mag, respectively,
using the same galaxy sample but
anchored to a UMa distance of 18.6 Mpc
and employing different correction methods for
the Galactic and internal extinction.
We also note that Eq.~\ref{eqtf} is
consistent with the relation found by \citet{sak00}
for a sample containing both field and cluster galaxies,
which demonstrates
that the relation does not vary
significantly with environment.

The TF parameters for Maffei~2 are given in Table~\ref{tabtf}.
Using our $I$-band TF relation (Eq.~\ref{eqtf}), we obtain
a distance to Maffei~2 of $\dismafii \pm \ddismafii$~Mpc
($\mu=\mumafii \pm \dmumafii$~mag).
The random uncertainty is computed from
\begin{equation}
\label{eqtferr}
(\delta M_I)^2 = [m \, \delta \log{2\vflat}]^2 + [\delta m (\log{2\vflat}-\langle\log{2\vflat}\rangle)]^2
+ \sigma_{M_I}^2/n
\end{equation}
where $\delta m$ is the standard error associated with the slope $m$
of the TF relation (Eq.~\ref{eqtf}),
$\delta \log{2\vflat}$ is the uncertainty in the estimate
of $\log{2\vflat}$ for Maffei~2,
$\sigma_{M_I}$ is the rms deviation in $M_I$ of the
$n=15$ galaxies in the V01 sample,
and $\langle\log{2\vflat}\rangle$ is the mean
value of $\log{2\vflat}$ for the V01 sample,
found to be 2.45.
The systematic uncertainty associated with the calibration
of the distance to the UMa cluster amounts to 0.13~mag.

V01 also measured total $\kp$ magnitudes
and report a $\kp$--band TF relation with
a slightly reduced scatter of 0.26 mag in $M_{\kp}$
as opposed to \rmstfi~mag in $M_I$.
Maffei~2 has been observed in $K_s$ by the
Two Micron All Sky Survey (2MASS),
and an estimate of its total magnitude, \kext, is available.
However, we find that the values of \kext\ reported by 2MASS
for the V01 sample differ considerably from
the total $\kp$ magnitudes measured by V01,
with a mean difference of $0.3\pm0.4$~mag
and the discrepancy reaching as high as 1.2 mag.
The large standard deviation in the residuals
reflects a significant difference in the measurement
techniques adopted by the two surveys.
The V01 total magnitude of a galaxy was obtained by integration
of the fit to the growth curve out to infinity,
while the 2MASS extrapolated magnitude was computed
by integration of the fit out to a finite radius
judged to contain the extent of the galaxy.
Hence, the V01 total magnitudes allow for a more
meaningful comparison of the integrated properties.
Further evidence of this is the significantly
larger scatter of \rmstfkext~mag found for the TF relation
constructed from the 2MASS \kext\ magnitudes
for the V01 sample.

In an attempt to derive a transformation from
2MASS to V01 magnitudes,
we investigated the magnitude residuals as
a function of various galaxy parameters.
The only dependency we observed is on
the apparent $\kp$-band
central disk surface brightness, $\mu_0(\kp)$,
from \citet{tul96},
in the sense that $\kext{-}\kp$ is larger
for the galaxies with low surface brightness
(ie., $\mu_0(\kp)>17 \rm \, mag \, arcsec^{-2}$).
This is illustrated in Fig.~\ref{figresid}.
Unfortunately, the large scatter in $\kext{-}\kp$
does not allow for a reliable estimate
of a $\kp$ magnitude for Maffei~2
from \kext.
The lack of a V01 measurement of the total $\kp$ magnitude
for Maffei~2 therefore precludes a reliable TF distance
in this bandpass,
so we adopt the $I$--band value of \dismafii~Mpc.

\subsection{The Distance to IC~342}
\label{s_dist_ic}

The small inclination of IC~342 makes it difficult to
determine an accurate distance from the TF relation.
However, recent observations
of Cepheids make possible a reliable determination
from the PL relation.
\citet{sch02} observed Cepheid variables in IC~342 in
Thuan-Gunn $r$ and $i$, deriving a distance modulus of $27.58 \pm 0.18$~mag
and extinction
$A_V = 2.01 \pm 0.32$~mag.  However, the adopted PL
relations for $r$ and $i$ were taken from \citet{hoe94},
who derived them from the LMC relations for Cousins $R$ and $I$
presented by \citet{mf91},
who in fact employed a different $I$-band PL relation
from that adopted in the \citet{fre01}.
Also, the transformation equations
employed to get the PL relations in $r$ and $i$ \citep{wad79}
were founded upon
Johnson $R$ and $I$ (hereafter $R_J$ and $I_J$), not Cousins 
$R$ and $I$ (hereafter $R_C$ and $I_C$). 
Thus, a re-examination of the Cepheid
distance is in order.

PL relations for the LMC in $R_J$ and $I_J$
have been derived by combining the \citet{fre01} relations
for $V$ and $I_C$ as revised in \S~\ref{s_zp},
the period-colour relation
for $V - R_C$ \citep{mf91}, and
transformation equations connecting $V - R_C$ to $V - R_J$ and
$V - I_C$ to $V - I_J$ \citep{bes79}.  In turn, revised
PL relations for $r$ and $i$ have been derived from those for $R_J$
and $I_J$ using the transformation equations of
\citet{wad79}.  On a
scale where the LMC is at a distance modulus of 18.50 mag,
\begin{eqnarray}
M_r & = & -3.002 (\log P - 1) - 4.104 \\
M_i & = & -3.012 (\log P - 1) - 4.159
\end{eqnarray}
where P is the period in days.
With the new PL relations, the mean apparent distance moduli in
$r$ and $i$ for the
Cepheids in IC~342 become $\mu_r = 29.100 \pm 0.045$~mag and
$\mu_i = 28.712 \pm 0.042$~mag, respectively, and the mean colour excess
becomes $E(r-i) = \mu_r - \mu_i = 0.388 \pm 0.036$~mag.  
The colour excess permits
a solution for the optical depth of dust localized in IC~342
(\tauex),
since the optical depth of Galactic dust at $1 \, \rm \mu m$
has been shown
to be $\taugal = \tauic \pm \dtauic$ (\S~\ref{s_ext}).
Knowledge of the optical
depth of both components then permits corrections for 
extinction in $r$ and $i$.  Calculations have been accomplished
with YES using the reddening
law of \citet{fi99} tuned to give $A_V / E(B-V) = 3.07$
for Vega \citep[see][]{mlm04}.

Adopting a heliocentric velocity
of $25 \, \kms$ \citep{bm99},
the observed mean colour excess of
the Cepheids gives 
$\tauex = 0.056 \hbox{--} 0.064$ for spectral types between
F8~I and G2~I, typical of luminous Cepheids.  Solving for the
Galactic and localized components of the extinction in $i$ 
over the same span of spectral types (the
K-correction is negligible), the true distance modulus $\mu^0$
ranges from 27.504 to 27.492 mag.  

The distance modulus can be corrected for metallicity using
observations of HII regions in IC~342
made by \citet{mrs85}. 
As was done for \citet{fre01} Cepheids,
measurements of $R_{23}$ across the disk were converted to
oxygen abundances using the
calibration of \citet{zar94}.  Based upon their
positions, the
appropriate mean metallicity to adopt for the Cepheids 
observed in IC~342 is $< \log Z > = 9.12$.  
With $\log Z_{LMC} = 8.50$ \citep{sak00}, the
corrected distance modulus is $\mu^0_Z = 27.643 \pm 0.120$~mag on the scale
where the LMC is at 18.50 mag.  The uncertainty is that due to
random errors, specifically the quadrature sum of the errors associated
with $E(r-i)$ and with the apparent distance
modulus in $i$. The result is 0.06 mag higher than
the value estimated by \citet{sch02}.  Note also that the
revised extinction of the Cepheids in $V$ is $2.10 \, \rm mag$
(Galactic plus localized), which is 
$0.09 \, \rm mag$ more than
estimated by \citet{sch02}.  Correcting to the maser zero-point,
the distance modulus for IC~342 becomes $\muic \pm \dmuic$~mag.

\subsection{The Distance to Maffei~1}
\label{s_maf1}

Recently, the extinction of Maffei~1 was found
from the colour-$\rm Mg_2$ relation to be $\taugal=1.691 \pm 0.066$
\citep{fmd03}.
The corresponding broadband extinctions were derived from
extinction
coefficients computed by an older version of YES than that employed
in this paper for Maffei~2 and IC~342.
In the updated version of YES, the accuracy of
extinction computations has increased owing to improvements
in the template SEDs \citep[see][]{mlm04}.
For consistency with the treatment of Maffei~2 and IC~342,
we have re-computed the extinction parameters of Maffei~1
using the same version of YES employed in this paper.
The revised extinction coefficients,
broadband extinctions and extinction-corrected magnitudes for Maffei~1
are given in Table~\ref{tabprop}.

\citet{fmd03} used the $I$-band FP
relation constructed
from ellipticals in the Coma cluster to determine the distance to Maffei~1.
The authors arrived at a distance of
$2.99 \pm 0.30$~Mpc, which is the
weighted mean of their FP estimate combined with
measurements from
the $D_n-\sigma$ relation of \citet{lfb88}
using photometry in $B$ and $\kp$.
The $D_n-\sigma$ distance estimates depend on the Hubble Constant.
To place \ho\ on the same extinction scale as that adopted
in this paper
would require a re-analysis of $all$ \citet{fre01} zero-points
(see \S~\ref{s_zp}).
Given that the two $D_n-\sigma$ distances to Maffei~1 are consistent
within errors with the FP distance derived for this galaxy,
here we adopt the FP distance alone in order to obtain a value
that is on the same extinction and distance scale as the
distances to Maffei~2 and IC~342 derived in this paper.
The FP estimate depends on the $I$-band extinction of Maffei~1
as well as
the adopted distance to the Coma cluster.
The modifications to the extinction of Maffei~1 and the
distance to the Coma cluster (see \S~\ref{s_zp}) result in
a revised distance to Maffei~1 of $\dismafi \pm \ddismafi$~Mpc
(corrected to the NGC~4258 maser zero-point).

\section{Implications for the IC~342/Maffei Group}
\label{implications}

The $I$-band imaging of \citet{bm99} revealed
a close grouping of galaxies apparently dominated
by the spirals Maffei~2 and IC~342 and the giant elliptical
Maffei~1.
Unfortunately, the gross uncertainties in the distances to
these three galaxies, owing to their heavy obscuration,
precluded a definitive statement
about their relative positions and luminosities.
The results of this paper now allow for a reliable
comparison of all three galaxies.

The revised properties of Maffei~1, Maffei~2 and IC~342 are
summarized in Table~\ref{tabprop}.
All values are now
anchored to the same distance and extinction scale.
The implications of our results for these
three galaxies are as follows:

1. The distances found in this study place the three
galaxies within \depgrp~Mpc in depth.
Their maximum spread in angle is $10^\circ$,
which corresponds to a metric distance of \spreadgrp~Mpc
at the average distance to the galaxies of $\disgrp \pm \ddisgrp$~Mpc.

2. Maffei~1, Maffei~2 and IC~342 have nearly identical
face-on luminosities, spanning a range of only \spreadmi~mag in $M_I$.
As predicted by \citet{spi73},
Maffei~2 suffers more Galactic extinction than Maffei~1
by nearly a full magnitude in $V$.
Although a magnitude fainter in $M_I$ than M31,
the three galaxies
clearly dominate the IC~342/Maffei group.
The next brightest member, Dwingeloo 1,
has a rotational velocity comparable to M33
and can be considered to be $\sim2$~mag fainter
than the giants in $I$.

3. Velocities in the Local Group reference frame, computed using
the solar apex vector of \citet{cv99},
are 311~\kms, 220~\kms\ and 255~\kms\ for Maffei~1,
Maffei~2 and IC~342, respectively.
These values significantly exceed the velocity dispersion
of 61~\kms\ for the Local Group \citep{cv99},
placing all three galaxies well beyond its dynamical range.

Previously,
the role that Maffei~1, Maffei~2 and IC~342
have played in the evolution of the Milky Way
and its neighbours has been a subject of controversy
due to their uncertain distances.
Early estimates of the radial velocities and distances
to Maffei~1 and IC~342 suggested that the galaxies could
have been in the vicinity of the Local Group within the
last 7 billion years, thereby affecting the early dynamical
evolution of the Local Group \citep{val93}.
Shortly following this analysis, \citet{kri95} used the angular
separations and radial velocities of 12 galaxies identified
as members of the IC~342/Maffei Group to argue that 
the evolution of the group must have been independent
of the Local Group.
More recently, distances to several dwarf galaxies in the
IC~342/Maffei group have been derived by \citet{kar03} via
the Tip of the Red Giant Branch (TRGB),
placing the dwarfs outside the early dynamical influence of the 
Local Group.
However,
past distance estimates to Maffei~2
were too uncertain to ascertain its
placement within the group or
its role as a dominant member.
The modern, homogeneous properties presented in this paper 
for all three galaxies
firmly establish
the existence of a compact group of galaxies dominated
by three giants at an average distance of \disgrp~Mpc from the Milky Way.
The mean velocity of the galaxies relative to the Local Group
is \vlggrp~\kms. Assuming a smooth Hubble flow with
$\ho = \hub$ (see \S~\ref{s_zp}),
the Hubble distance to the IC~342/Maffei group is \dishubgrp~Mpc.
This is in close agreement with the mean distance of
$\disgrp \pm \ddisgrp$~Mpc
found in this study. It is therefore highly unlikely
that the dominant members of the IC~342/Maffei group have
interacted with the Local Group since the Big Bang.
This conclusion lends credence to the timing model of \citet{kw59},
which is based on the simple evolutionary scenario of a
two-body system involving M31 and the Milky Way with
negligible perturbation from other galaxies.
The model provides an estimate for the age of the Universe
and mass of the Local Group, in particular offering strong
evidence that most of the mass in the Local Group is dark.
The predictions of the model are therefore strengthened
by the results of this study.

\section{Summary}
\label{summary}

We have constructed a correlation between dust and gas in
spiral galaxies as a tool for determining the Galactic
extinction of a heavily obscured spiral.
The correlation
can be used to estimate the extragalactic extinction
of constituent HII regions from measurements of the
column density of hydrogen in the heavily
obscured galaxy.
The spiral's Galactic extinction can then be determined
by taking the average difference between the extragalactic
extinction and the total extinction of each HII region
found from observations
of the Balmer decrement.

We have used the dust--gas correlation to measure
the Galactic extinction of the heavily obscured spirals
Maffei~2 and IC~342, for which
we find values of $\taugal=\taumafii \pm \dtaumafii$
and $\taugal=\taucic \pm \dtaucic$,
respectively,
where \taugal\ is the optical depth of Galactic dust at 1~$\mu$m.
The Galactic extinction of Maffei~2 is the most
accurate measurement to date,
while the result for IC~342 is consistent
with the optical depth of $\tausfdic \pm \dtausfdic$
obtained from the reddening maps of \citet{sfd98}. 
We therefore adopt the weighted mean of the two values for IC~342,
which yields $\taugal=\tauic \pm \dtauic$.
Thus, $A_V=\avmafii \pm \davmafii$~mag for Maffei~2
and $A_V=\avic \pm \davic$~mag for IC~342.
The reddening coefficients used in this analysis
were computed by the York Extinction Solver (YES),
which accounts for the colour-dependent shifts
in the effective wavelengths of broadband filters.

To facilitate distance determinations, we have re-analyzed
the zero-points of the extragalactic distance scale using the approach
to extinction adopted for the galaxies in the IC~342/Maffei group.

To determine the distance to Maffei~2,
we have formulated the Tully-Fisher (TF) relation in $I$
using rotation curves instead of HI line widths as the
luminosity diagnostic,
as line widths for nearby galaxies like Maffei~2 can be unreliable
due to contamination by Galactic HI.
The TF distance to Maffei~2
is found to be $\dismafii \pm \ddismafii$~Mpc,
where the zero-point is set by the maser distance to
NGC~4258.
Using the extinction treatment and distance scale adopted in this paper,
we re-derive the Cepheid distance to IC~342
\citep{sch02} and the Fundamental Plane distance to Maffei~1 \citep{fmd03},
obtaining $\disic \pm \ddisic$~Mpc
and $\dismafi \pm \ddismafi$~Mpc, respectively.
The revised distances and magnitudes reveal
that the IC~342/Maffei group is dominated by three giant
galaxies with nearly identical luminosities.
Further, the average distance to the three galaxies is in excellent
agreement with expectations for the Hubble Flow, making it highly unlikely
that these galaxies interacted with the Local Group since the
Big Bang.

\acknowledgments

RLF and MLM gratefully acknowledge the
continuing support of the Natural Sciences and Engineering Research
Council of Canada and the Ontario Graduate Scholarship Program.
MLM is also grateful to S. Sakai for very helpful communications
and to R. Ross for performing initial data reductions.
HL and MGR thank NOAO for the use of their facilities.
MGR gratefully acknowledges the financial support of grants
from CONACYT (CONACYT 43121) and UNAM (DGAPA IN 112103, IN 108406-2
and IN 108506-2).
This publication makes use of data products from the Two Micron
All Sky Survey, which is a joint project of the University of
Massachusetts and the Infrared Processing and Analysis
Center/California Institute of Technology, funded by the
National Aeronautics and Space Administration and the National
Science Foundation.
Some data were accessed as Guest User, Canadian Astronomy Data Centre,
which is operated by the Dominion Astrophysical Observatory
for the National Research Council of Canada's
Herzberg Institute of Astrophysics.
In addition, this research has made use of the NASA/IPAC
Extragalactic Database (NED) which is operated by the
Jet Propulsion Laboratory, California Institute of Technology,
under contract with the National Aeronautics and Space Administration.


\clearpage
\begin{deluxetable}{cccccc}
\tabletypesize{\scriptsize}
\tablecolumns{6}
\tablewidth{0pt}
\tablecaption{Measurements of HII Regions in Maffei~2 and IC~342\label{tablines}}
\tablehead{
\colhead{GALAXY} & \multicolumn{2}{c}{Maffei~2} & \colhead{} & \multicolumn{2}{c}{IC~342} \\
\cline{2-3} \cline{5-6} \\
\colhead{HII REGION\tablenotemark{a}} & \colhead{[-0095,-0045]} & \colhead{[-0091,-0065]} & \colhead{} & \colhead{[+0055,-0333]} & \colhead{[+0072,-0292]}\\
\cline{1-6} \\
\colhead{LINE} & \multicolumn{5}{c}{$f/f_{\hb}$\tablenotemark{b}}
}
\startdata
\hg & \nodata & \nodata & & $27.0\pm3.4$ & \nodata \\
\hb & $100\pm28$ & $100\pm31$ & & $100\pm5.2$ & $100\pm8.2$ \\
\oiiia & \nodata & \nodata & & $29.7\pm2.6$ & $10.2\pm3.7$ \\
\oiiib & \nodata & \nodata & & $88.6\pm3.2$ & $19.2\pm4.4$ \\
HeI $\lambda5876$ & \nodata & \nodata & & $19.4\pm3.1$ & \nodata \\
$\rm [OI] \lambda6300$ & \nodata & \nodata & & $18.8\pm1.7$ & \nodata \\
$\rm [NII] \lambda6548$ & $347\pm165$ & $474\pm92$ & & \nodata & \nodata \\
\ha & $4376\pm201$ & $5867\pm113$ & & $637\pm34$ & $668\pm33$ \\
$\rm [NII] \lambda6583$ & $1613\pm172$ & $1601\pm94$ & & $195\pm28$ & $195\pm27$ \\
HeI $\lambda6678$ & \nodata & \nodata & & $11.8\pm2.1$ & \nodata \\
$\rm [SII] \lambda6716$ & $606\pm39$ & $719\pm94$ & & $61.1\pm2.3$ & $71.4\pm7.0$ \\
$\rm [SII] \lambda6731$ & $448\pm37$ & $565\pm90$ & & $41.8\pm2.2$ & $45.5\pm6.5$ \\
$\rm [ArIII] \lambda7136$ & \nodata & \nodata & & $18.4\pm2.5$ & \nodata \\
\tableline
$f_{\hb} \rm \, (10^{-16} \, ergs \, s^{-1} \, cm^{-2})$ & $1.09\pm0.30$ & $1.06\pm0.33$ & & $45.1\pm2.4$ & $22.4\pm1.8$ \\
$W_{\hb}$~(\AA)\tablenotemark{c} & $16.5\pm5.2$ & $6.6\pm2.1$ & & $442\pm224$ & $112\pm22$ \\
\enddata
\tablenotetext{a}
{
Eastward and northward offset of the HII region from the galaxy nucleus as projected on the sky, in arcsec.
}
\tablenotetext{b}
{
Uncorrected line flux relative to \hb\ scaled to $f_{\hb}=100$.
The errors in the line ratios do not include the error in $f_{\hb}$.
}
\tablenotetext{c}
{
Equivalent width of \hb\ emission.
(For typical HII regions, the equivalent width of \ha\ emission is so strong that the correction to the \ha\ flux is negligible.)
}
\end{deluxetable}

\clearpage
\thispagestyle{empty}
\begin{deluxetable}{cccccccccccccc}
\rotate
\tabletypesize{\scriptsize}
\renewcommand{\tabcolsep}{1mm}
\tablecolumns{14}
\tablewidth{0pt}
\tablecaption{Properties of HII Regions in Maffei~2 and IC~342\label{tabhiiex}}
\tablehead{
\colhead{HII REGION} & \colhead{$r$} & \colhead{$W_{\hb}$} & \colhead{$\fab$} & \colhead{$\ox$} & \colhead{T} & \colhead{$(\fab)^0$} & \colhead{$\tautot$} & \colhead{$\nhi$} & \colhead{$X$} & \colhead{$\nhmol$} & \colhead{$\nh$} & \colhead{$\tauex$} & \colhead{REF} \\
\colhead{} & \colhead{$(\arcsec)$} & \colhead{(\AA)} & \colhead{} & \colhead{} & \colhead{(K)} & \colhead{} & \colhead{} & \colhead{$\rm (10^{20} \, cm^{-2})$} & \colhead{$\rm (10^{20} \, cm^{-2} (\kms)^{-1})$} & \colhead{$\rm (10^{20} \, cm^{-2})$} & \colhead{$\rm (10^{20} \, cm^{-2})$} & \colhead{} & \colhead{} \\
\colhead{(1)} & \colhead{(2)} & \colhead{(3)} & \colhead{(4)} & \colhead{(5)} & \colhead{(6)} & \colhead{(7)} & \colhead{(8)} & \colhead{(9)} & \colhead{(10)} & \colhead{(11)} & \colhead{(12)} & \colhead{(13)} & \colhead{(14)} \\
} \startdata
\cutinhead{Maffei2}$[-0091,-0065]$ & 168 & 6.6 & $45.62\pm0.14$ & -2.98\tablenotemark{\dag} & 4,572 & 3.07 & $2.46\pm0.30$ & $18.13\pm2.42$ & 2.24 & $29.12\pm8.08$ & $38.19\pm8.17$ & $0.38\pm0.09$ & \tablenotemark{a} \\
$[-0095,-0045]$ & 187 & 16.5 & $39.25\pm0.13$ & -3.00\tablenotemark{\dag} & 4,748 & 3.06 & $2.33\pm0.27$ & $18.90\pm2.64$ & 2.33 & $26.34\pm7.29$ & $35.79\pm7.41$ & $0.37\pm0.08$ & \tablenotemark{a} \\
\cutinhead{IC342}$[-0521,-0166]$ & 564 & 197.0 & $10.54\pm0.04$ & -3.76 & 10,330 & 2.86 & $1.19\pm0.09$ & $9.33\pm1.72$ & 7.61 & $0.00\pm0.00$ & $4.66\pm0.86$ & $0.19\pm0.07$ & MR85 \\
$[-0077,-0288]$ & 303 & 64.3 & $7.64\pm0.04$ & -3.01 & 4,835 & 3.05 & $0.84\pm0.09$ & $8.20\pm1.10$ & 2.37 & $7.34\pm2.02$ & $11.44\pm2.10$ & $0.23\pm0.16$ & MR85 \\
$[+0832,-0158]$ & 916 & 53.9 & $6.46\pm0.04$ & -3.76 & 10,390 & 2.86 & $0.74\pm0.09$ & $7.30\pm1.83$ & 7.71 & $0.00\pm0.00$ & $3.65\pm0.91$ & $0.18\pm0.08$ & MR85 \\
$[-0150,-0031]$ & 160 & \nodata & $7.68\pm0.05$ & -3.24\tablenotemark{\dag} & 6,527 & 2.97 & $0.87\pm0.10$ & $4.31\pm0.46$ & 3.40 & $23.15\pm6.28$ & $25.31\pm6.28$ & $0.31\pm0.48$ & BK82 \\
$[+0055,-0333]$ & 357 & 442.4 & $6.34\pm0.05$ & -3.33\tablenotemark{\dag} & 7,147 & 2.94 & $0.70\pm0.11$ & $8.69\pm1.26$ & 3.87 & $6.37\pm1.72$ & $10.71\pm1.83$ & $0.22\pm0.14$ & \tablenotemark{a} \\
$[+0072,-0292]$ & 321 & 112.0 & $6.57\pm0.06$ & -3.31\tablenotemark{\dag} & 7,032 & 2.95 & $0.73\pm0.13$ & $8.36\pm1.15$ & 3.78 & $8.52\pm2.30$ & $12.70\pm2.38$ & $0.24\pm0.18$ & \tablenotemark{a} \\
\enddata
\tablecomments{
(1) Eastward and northward offset of the HII region from the galaxy nucleus as projected on the sky, in arcsec.
(2) De-projected galactocentric radius of the HII region
using $i$ and $\phi$ from Table~\ref{tabspi}.
(3) Equivalent width of the $\hb$ emission line, where available.
(4) Observed $\hab$ flux ratio, corrected for underlying stellar
absorption but not for Galactic reddening, assuming an equivalent
width for absorption of 1.9~\AA \citep[see][]{mrs85}.
(5) Oxygen abundance of the HII region derived from the
forbidden oxygen lines (see \S~\ref{s_extot}).
(6) Equilibrium temperature of the HII region derived from $\ox$
(see \S~\ref{s_extot}).
(7) Intrinsic Balmer flux ratio based on the HII region's
equilibrium temperature and an electron density of 100 $\rm cm^{-3}$.
(8) Total optical depth at 1~$\mu$m of the HII region derived from the observed
and intrinsic Balmer decrements (Eq.~\ref{eqtau}).
(9) Annular averaged column density of neutral hydrogen within the
annulus at $r$.
(10) Conversion factor for CO intensity to $\hmol$ column density
($X = \nhmol/\ico$).
(11) Annular averaged column density of molecular hydrogen within the
annulus at $r$.
(12) Half of the column density of hydrogen in both atomic and molecular form
at $r$.
(13) Extragalactic optical depth at 1~$\mu$m of the HII region,
i.e., the difference between the total optical depth $\tautot$
and the Galactic optical depth $\taugal$.
For Maffei~2 and IC~342, \tauex\ is computed from Eq.~\ref{eqdust}.
(14) Source of the HII region spectrum.
Abbreviations are cross-referenced in the bibliography.
}
\tablenotetext{\dag}
{Indicates the absence of oxygen line data,
resulting in the use of Eq.~\ref{eqtsbc} or Eq.~\ref{eqtscd}
to determine the temperature, and hence $\ox$.}
\tablenotetext{a}
{This study}
\end{deluxetable}

\clearpage
\begin{deluxetable}{lcccccccccll}
\tabletypesize{\scriptsize}
\tablewidth{0pt}
\tablecaption{Properties of the Spirals Hosting HII Regions\label{tabspi}}
\tablehead{
\colhead{GALAXY} & \colhead{T} & \colhead{RA(2000)} & \colhead{DEC(2000)} & \colhead{$v_\odot$} & \colhead{$i$} & \colhead{$\phi$} & \colhead{$r_0$} & \colhead{$E(B-V)$} & \colhead{$\taugal$} & \colhead{HI REF} & \colhead{CO REF} \\
\colhead{} & \colhead{} & \colhead{(h:m:s)} & \colhead{(d:m:s)} & \colhead{$(\kms)$} & \colhead{$(\circ)$} & \colhead{$(\circ)$} & \colhead{$(\arcsec)$} & \colhead{(mag)} & \colhead{} & \colhead{} & \colhead{} \\
\colhead{(1)} & \colhead{(2)} & \colhead{(3)} & \colhead{(4)} & \colhead{(5)} & \colhead{(6)} & \colhead{(7)} & \colhead{(8)} & \colhead{(9)} & \colhead{(10)} & \colhead{(11)} & \colhead{(12)}
}
\startdata
NGC2903 & 4 & 09:32:09.7 & +21:30:02 & 560 & 60 & 29 & 58.2 & 0.031 & 0.035 & WV86 & YX95 \\
NGC4303 & 4 & 12:21:54.7 & +04:28:20 & 1,568 & 25 & 7 & 38.4 & 0.022 & 0.025 & W88 & KY88 \\
NGC4321 & 4 & 12:22:55.2 & +15:49:23 & 1,574 & 30 & 153 & 63.6 & 0.026 & 0.030 & W88 & KY88 \\
NGC4501 & 3 & 12:31:59.6 & +14:25:17 & 2,276 & 58 & 140 & 49.7 & 0.038 & 0.043 & W88 & YX95 \\
NGC5055 & 4 & 13:15:49.3 & +42:02:06 & 500 & 55 & 99 & 102.6 & 0.018 & 0.020 & WV86 & YX95 \\
NGC5194 & 4 & 13:29:53.3 & +47:11:48 & 464 & 20 & 170 & 111.6 & 0.035 & 0.040 & TA91 & YX95 \\
\cline{1-12} \\
Maffei2 & 4 & 02:41:54.6\tablenotemark{a} & +59:36:11\tablenotemark{a} & -23 & 67 & 206 & 118.9 & \nodata & \nodata & HT96 & MW04 \\
\cline{1-12} \\
NGC598 & 6 & 01:33:50.9 & +30:39:37 & -181 & 57 & 22 & 494.4 & 0.042 & 0.048 & N80a & YX95 \\
NGC2403 & 6 & 07:36:54.5 & +65:35:58 & 125 & 60 & 125 & 163.2 & 0.040 & 0.045 & WV86 & YX95 \\
NGC4654 & 6 & 12:43:56.6 & +13:07:33 & 1,030 & 52 & 128 & 39.7 & 0.026 & 0.030 & W88 & YX95 \\
NGC5457 & 6 & 14:03:12.5 & +54:20:55 & 243 & 18 & 39 & 126.6 & 0.009 & 0.010 & BG81 & KS91 \\
\cline{1-12} \\
IC342 & 6 & 03:46:49.7\tablenotemark{a} & +68:05:45\tablenotemark{a} & 25\tablenotemark{b} & 25\tablenotemark{b} & 39\tablenotemark{b} & 209.3 & \nodata & \nodata & N80b & YX95
\enddata
\tablecomments{
(1) Name of galaxy.
(2) Morphological stage from RC3.
(3) Right ascension (from RC3 unless otherwise noted).
(4) Declination (from RC3 unless otherwise noted).
(5) Heliocentric radial velocity from the HI reference.
(6) Inclination angle from the HI reference.
(7) Position angle from the HI reference.
(8) Disk scale length (the radius at which the surface brightness
in $B$ drops to $1/e$ of its central value) from \citealt{sk96}
(NGC4501 and NGC4654), \citealt{bm99} (Maffei~2 and IC~342)
and \citealt{zar94} for all other galaxies.
(9) Galactic reddening from \citealt{sfd98}.
(10) Optical depth of Galactic dust at 1~$\mu$m calculated from $E(B-V)$ (see \S~\ref{s_gasdust}).
(11) Source of the HI map of the galaxy. Abbreviations are cross-referenced in the bibliography.
(12) Source of the CO distribution of the galaxy. Abbreviations are cross-referenced in the bibliography.
}
\tablenotetext{a}
{\citealt{bm99}}
\tablenotetext{b}
{\citealt{new80a}}
\end{deluxetable}

\clearpage
\begin{deluxetable}{cccccccccccccc}
\rotate
\tabletypesize{\scriptsize}
\renewcommand{\tabcolsep}{1mm}
\tablecolumns{14}
\tablewidth{0pt}
\tablecaption{Properties of the HII Region Sample\label{tabhii}}
\tablehead{
\colhead{HII REGION} & \colhead{$r$} & \colhead{$W_{\hb}$} & \colhead{$\fab$} & \colhead{$\ox$} & \colhead{T} & \colhead{$(\fab)^0$} & \colhead{$\tautot$} & \colhead{$\nhi$} & \colhead{$X$} & \colhead{$\nhmol$} & \colhead{$\nh$} & \colhead{$\tauex$} & \colhead{REF} \\
\colhead{} & \colhead{$(\arcsec)$} & \colhead{(\AA)} & \colhead{} & \colhead{} & \colhead{(K)} & \colhead{} & \colhead{} & \colhead{$\rm (10^{20} \, cm^{-2})$} & \colhead{$\rm (10^{20} \, cm^{-2} (\kms)^{-1})$} & \colhead{$\rm (10^{20} \, cm^{-2})$} & \colhead{$\rm (10^{20} \, cm^{-2})$} & \colhead{} & \colhead{} \\
\colhead{(1)} & \colhead{(2)} & \colhead{(3)} & \colhead{(4)} & \colhead{(5)} & \colhead{(6)} & \colhead{(7)} & \colhead{(8)} & \colhead{(9)} & \colhead{(10)} & \colhead{(11)} & \colhead{(12)} & \colhead{(13)} & \colhead{(14)} \\
} \startdata
\cutinhead{NGC2903}$[+0024,+0047]$ & 53 & 30.8 & 3.99 & -2.88 & 3,877 & 3.13 & 0.22 & 14.26 & 1.94 & 11.92 & 38.10 & 0.19 & MR85 \\
$[-0003,-0070]$ & 89 & 27.7 & 5.61 & -2.99 & 4,641 & 3.07 & 0.55 & 15.90 & 2.28 & 10.42 & 36.74 & 0.52 & MR85 \\
$[-0003,+0069]$ & 93 & 27.6 & 6.56 & -2.92 & 4,134 & 3.11 & 0.68 & 15.78 & 2.04 & 9.24 & 34.26 & 0.65 & MR85 \\
$[+0039,-0078]$ & 152 & 73.0 & 4.36 & -3.21 & 6,272 & 2.98 & 0.35 & 11.14 & 3.22 & 13.94 & 39.02 & 0.31 & MR85 \\
$[+0047,-0072]$ & 157 & 31.1 & 3.12 & -3.01 & 4,820 & 3.05 & 0.02 & 10.51 & 2.36 & 10.24 & 30.99 & -0.02 & MR85 \\
$[-0021,+0105]$ & 161 & 41.8 & 4.55 & -3.21 & 6,280 & 2.98 & 0.39 & 10.07 & 3.22 & 13.95 & 37.97 & 0.35 & MR85 \\
\cutinhead{NGC4303}$[-0001,+0045]$ & 45 & 33.0 & 3.84 & -2.93 & 4,255 & 3.10 & 0.20 & 11.67 & 2.10 & 19.97 & 51.61 & 0.16 & SS91 \\
$[-0013,-0044]$ & 46 & 36.0 & 3.40 & -3.09 & 5,393 & 3.02 & 0.11 & 11.62 & 2.67 & 24.63 & 60.88 & 0.07 & SS91 \\
$[-0014,+0048]$ & 51 & 160.0 & 5.18 & -2.98 & 4,569 & 3.07 & 0.48 & 11.36 & 2.24 & 17.49 & 46.34 & 0.44 & SS91 \\
$[+0046,+0006]$ & 51 & 19.0 & 3.65 & -2.93 & 4,185 & 3.10 & 0.15 & 5.76 & 2.07 & 16.09 & 37.94 & 0.11 & SS91 \\
$[+0032,-0040]$ & 54 & 64.0 & 3.68 & -3.09 & 5,422 & 3.02 & 0.18 & 6.55 & 2.69 & 18.78 & 44.11 & 0.15 & SS91 \\
$[+0022,+0067]$ & 71 & 21.0 & 3.34 & -2.94 & 4,295 & 3.09 & 0.07 & 10.07 & 2.12 & 8.25 & 26.57 & 0.03 & SS91 \\
$[+0005,-0073]$ & 73 & 46.0 & 4.12 & -3.01 & 4,824 & 3.05 & 0.27 & 10.56 & 2.37 & 8.42 & 27.40 & 0.24 & SS91 \\
$[-0049,-0094]$ & 107 & 207.0 & 3.27 & -3.31 & 7,002 & 2.95 & 0.09 & 9.17 & 3.76 & 4.11 & 17.39 & 0.06 & SS91 \\
$[-0110,+0075]$ & 144 & 117.0 & 3.12 & -3.39 & 7,647 & 2.92 & 0.06 & 7.09 & 4.31 & 1.32 & 9.73 & 0.02 & SS91 \\
\cutinhead{NGC4321}$[-0051,+0009]$ & 57 & 21.2 & 5.87 & -2.93 & 4,196 & 3.10 & 0.58 & 5.87 & 2.07 & 12.81 & 31.49 & 0.55 & MR85 \\
$[-0001,-0066]$ & 68 & 39.0 & 4.59 & -2.97 & 4,481 & 3.08 & 0.36 & 6.13 & 2.20 & 10.63 & 27.39 & 0.33 & SS91 \\
$[+0032,-0074]$ & 81 & 33.4 & 5.55 & -2.92 & 4,135 & 3.11 & 0.53 & 5.71 & 2.04 & 7.72 & 21.15 & 0.49 & MR85 \\
$[+0013,+0102]$ & 108 & 62.0 & 4.06 & -2.98 & 4,625 & 3.07 & 0.25 & 7.06 & 2.27 & 5.08 & 17.22 & 0.22 & SS91 \\
$[-0114,+0010]$ & 127 & 43.5 & 4.67 & -3.01 & 4,795 & 3.06 & 0.39 & 6.95 & 2.35 & 3.68 & 14.31 & 0.35 & MR85 \\
$[-0131,-0027]$ & 153 & 137.0 & 3.77 & -3.11 & 5,516 & 3.02 & 0.20 & 5.80 & 2.74 & 2.66 & 11.12 & 0.17 & SS91 \\
$[-0032,+0147]$ & 152 & 33.0 & 4.70 & -3.23 & 6,404 & 2.97 & 0.42 & 5.87 & 3.31 & 3.28 & 12.43 & 0.38 & SS91 \\
$[+0029,+0146]$ & 158 & 68.0 & 4.22 & -3.08 & 5,346 & 3.02 & 0.30 & 9.29 & 2.64 & 2.34 & 13.97 & 0.27 & SS91 \\
$[+0034,+0145]$ & 159 & 51.6 & 4.50 & -3.04 & 5,003 & 3.04 & 0.36 & 9.00 & 2.46 & 2.14 & 13.28 & 0.32 & MR85 \\
\cutinhead{NGC4501}$[+0018,+0012]$ & 41 & 67.0 & 4.79 & -2.89\tablenotemark{\dag} & 3,911 & 3.13 & 0.39 & 9.83 & 1.95 & 15.46 & 40.75 & 0.35 & SK96 \\
$[-0043,-0005]$ & 72 & 39.0 & 5.16 & -2.98\tablenotemark{\dag} & 4,616 & 3.07 & 0.47 & 6.81 & 2.26 & 8.46 & 23.73 & 0.44 & SK96 \\
$[-0027,+0062]$ & 74 & 12.0 & 3.31 & -2.88 & 3,880 & 3.13 & 0.05 & 6.92 & 1.94 & 6.90 & 20.72 & 0.02 & SK96 \\
$[-0094,+0049]$ & 124 & 70.0 & 3.31 & -3.14\tablenotemark{\dag} & 5,776 & 3.00 & 0.09 & 5.55 & 2.90 & 3.14 & 11.83 & 0.05 & SK96 \\
$[-0068,+0093]$ & 116 & 26.0 & 3.84 & -3.12\tablenotemark{\dag} & 5,589 & 3.01 & 0.22 & 6.28 & 2.78 & 3.68 & 13.64 & 0.19 & SK96 \\
\cutinhead{NGC5055}$[-0089,-0010]$ & 96 & 16.3 & 5.06 & -2.92 & 4,164 & 3.11 & 0.45 & 15.53 & 2.06 & 9.95 & 35.43 & 0.41 & MR85 \\
$[-0165,+0034]$ & 169 & 18.7 & 5.20 & -2.99 & 4,650 & 3.07 & 0.48 & 15.78 & 2.28 & 11.03 & 37.84 & 0.45 & MR85 \\
$[+0050,-0088]$ & 152 & 52.2 & 6.13 & -3.10 & 5,442 & 3.02 & 0.65 & 16.16 & 2.70 & 13.05 & 42.26 & 0.61 & MR85 \\
$[-0270,+0068]$ & 281 & 118.6 & 3.36 & -3.24 & 6,524 & 2.97 & 0.11 & 8.07 & 3.39 & 16.42 & 40.91 & 0.08 & MR85 \\
\cutinhead{NGC5194}$[-0007,+0061]$ & 61 & 50.4 & 4.79 & -2.89 & 3,913 & 3.13 & 0.39 & 5.07 & 1.95 & 37.59 & 80.25 & 0.35 & MR85 \\
$[-0087,-0082]$ & 125 & 51.9 & 3.84 & -2.90 & 3,996 & 3.12 & 0.19 & 4.87 & 1.99 & 9.32 & 23.51 & 0.15 & MR85 \\
\cutinhead{NGC598}$[-0606,-1708]$ & 1,816 & 4.3 & 2.67 & -3.20 & 6,184 & 2.98 & -0.10 & 9.42 & 3.16 & 0.00 & 9.42 & -0.15 & MR85 \\
$[-0499,-0054]$ & 846 & 20.5 & 3.95 & -3.24 & 6,518 & 2.97 & 0.26 & 18.01 & 3.39 & 0.00 & 18.01 & 0.21 & MR85 \\
$[-0185,+0163]$ & 435 & 203.0 & 4.23 & -3.28 & 6,773 & 2.96 & 0.33 & 19.03 & 3.58 & 0.00 & 19.03 & 0.28 & MR85 \\
$[+0034,-0037]$ & 86 & 28.7 & 4.05 & -3.14 & 5,739 & 3.01 & 0.27 & 14.93 & 2.87 & 7.56 & 30.05 & 0.22 & MR85 \\
$[+0140,-0042]$ & 268 & \nodata & 4.84 & -3.27 & 6,752 & 2.96 & 0.45 & 15.61 & 3.56 & 0.00 & 15.61 & 0.40 & KA81 \\
$[+0140,+0340]$ & 368 & 117.0 & 3.49 & -3.17 & 5,998 & 2.99 & 0.14 & 16.09 & 3.04 & 0.00 & 16.09 & 0.09 & VP88 \\
$[-0210,+0123]$ & 444 & \nodata & 4.76 & -3.37 & 7,488 & 2.93 & 0.44 & 19.48 & 4.16 & 0.00 & 19.48 & 0.40 & KA81 \\
$[+0308,-0132]$ & 615 & 107.0 & 2.98 & -3.21 & 6,292 & 2.98 & 0.00 & 18.26 & 3.23 & 0.00 & 18.26 & -0.05 & VP88 \\
$[+0139,+0736]$ & 782 & \nodata & 4.31 & -3.55 & 8,814 & 2.89 & 0.36 & 15.19 & 5.52 & 0.00 & 15.19 & 0.32 & KA81 \\
$[+0540,+0458]$ & 871 & 86.0 & 3.78 & -3.33 & 7,190 & 2.94 & 0.23 & 20.21 & 3.91 & 0.00 & 20.21 & 0.18 & VP88 \\
$[-0520,+0345]$ & 1,129 & 20.0 & 2.84 & -3.42 & 7,830 & 2.92 & -0.02 & 19.87 & 4.48 & 0.00 & 19.87 & -0.07 & VP88 \\
$[+0526,+1245]$ & 1,352 & \nodata & 3.07 & -3.39 & 7,642 & 2.92 & 0.04 & 17.83 & 4.30 & 0.00 & 17.83 & 0.00 & KA81 \\
$[-0467,+0813]$ & 1,473 & \nodata & 4.48 & -3.63 & 9,425 & 2.88 & 0.40 & 19.87 & 6.28 & 0.00 & 19.87 & 0.36 & KA81 \\
$[-0857,-0039]$ & 1,476 & 97.0 & 3.20 & -3.46 & 8,163 & 2.91 & 0.09 & 19.93 & 4.80 & 0.00 & 19.93 & 0.04 & VP88 \\
$[-0454,+1029]$ & 1,675 & \nodata & 4.48 & -3.61 & 9,245 & 2.88 & 0.40 & 15.22 & 6.04 & 0.00 & 15.22 & 0.35 & KA81 \\
\cutinhead{NGC2403}$[-0494,+0137]$ & 592 & 157.0 & 3.49 & -3.55 & 8,797 & 2.89 & 0.17 & 11.05 & 5.50 & 0.02 & 11.09 & 0.13 & MR85 \\
$[-0133,-0146]$ & 393 & 155.9 & 3.40 & -3.43 & 7,906 & 2.92 & 0.14 & 15.67 & 4.55 & 0.09 & 15.85 & 0.09 & MR85 \\
$[+0010,+0032]$ & 65 & 83.1 & 3.76 & -3.27 & 6,747 & 2.96 & 0.22 & 20.30 & 3.56 & 1.55 & 23.40 & 0.17 & MR85 \\
$[+0045,+0069]$ & 165 & 180.5 & 4.26 & -3.40 & 7,678 & 2.92 & 0.34 & 21.17 & 4.33 & 0.56 & 22.29 & 0.30 & MR85 \\
$[+0063,-0049]$ & 80 & 72.0 & 4.06 & -3.16 & 5,938 & 3.00 & 0.28 & 20.24 & 3.00 & 0.96 & 22.16 & 0.23 & MR85 \\
$[+0165,+0136]$ & 416 & 213.0 & 3.60 & -3.59 & 9,087 & 2.89 & 0.20 & 15.07 & 5.84 & 0.09 & 15.25 & 0.16 & MR85 \\
\cutinhead{NGC5457}$[-0376,-0063]$ & 390 & 61.9 & 3.40 & -3.30 & 6,945 & 2.95 & 0.13 & 8.63 & 3.71 & 1.08 & 10.79 & 0.12 & MR85 \\
$[-0243,+0163]$ & 308 & 169.3 & 4.52 & -3.40 & 7,686 & 2.92 & 0.40 & 9.96 & 4.34 & 2.55 & 15.06 & 0.39 & MR85 \\
$[+0098,+0272]$ & 291 & 68.2 & 3.42 & -3.32 & 7,088 & 2.95 & 0.14 & 10.17 & 3.82 & 2.60 & 15.37 & 0.13 & MR85 \\
$[+0223,-0127]$ & 269 & 234.3 & 3.12 & -3.28 & 6,807 & 2.96 & 0.05 & 10.08 & 3.60 & 2.93 & 15.94 & 0.04 & MR85 \\
$[+0252,-0107]$ & 287 & 249.4 & 4.51 & -3.56 & 8,879 & 2.89 & 0.40 & 10.22 & 5.59 & 3.93 & 18.08 & 0.39 & MR85 \\
$[+0666,+0172]$ & 701 & 213.6 & 3.35 & -3.67\tablenotemark{\dag} & 9,658 & 2.87 & 0.14 & 5.91 & 6.60 & 0.13 & 6.17 & 0.13 & MR85 \\
$[+0167,+0012]$ & 172 & 162.2 & 4.14 & -3.29\tablenotemark{\dag} & 6,918 & 2.95 & 0.31 & 8.33 & 3.69 & 6.91 & 22.15 & 0.30 & RP82 \\
$[-0068,-0090]$ & 113 & \nodata & 4.84 & -3.25\tablenotemark{\dag} & 6,611 & 2.97 & 0.45 & 7.34 & 3.46 & 10.74 & 28.82 & 0.44 & RP82 \\
$[+0145,-0140]$ & 212 & \nodata & 4.41 & -3.32\tablenotemark{\dag} & 7,124 & 2.94 & 0.37 & 9.68 & 3.85 & 5.14 & 19.96 & 0.36 & RP82 \\
$[+0669,+0174]$ & 704 & 173.8 & 3.84 & -3.67\tablenotemark{\dag} & 9,676 & 2.87 & 0.26 & 5.86 & 6.62 & 0.13 & 6.12 & 0.25 & RP82 \\
$[+0250,-0113]$ & 288 & 195.0 & 5.03 & -3.38\tablenotemark{\dag} & 7,517 & 2.93 & 0.49 & 10.21 & 4.19 & 2.92 & 16.05 & 0.48 & RP82 \\
\cutinhead{NGC4654}$[-0068,+0033]$ & 78 & 83.0 & 4.14 & -3.28 & 6,800 & 2.96 & 0.31 & 20.00 & 3.60 & 4.09 & 28.18 & 0.28 & SK96 \\
$[-0034,-0056]$ & 106 & 100.0 & 3.08 & -3.31 & 7,054 & 2.95 & 0.04 & 12.68 & 3.80 & 1.47 & 15.62 & 0.01 & SK96 \\
$[-0055,+0051]$ & 75 & 127.0 & 3.57 & -3.36 & 7,364 & 2.93 & 0.18 & 19.72 & 4.06 & 5.12 & 29.96 & 0.15 & SK96 \\
$[-0042,+0035]$ & 55 & 59.0 & 3.57 & -3.11 & 5,554 & 3.01 & 0.15 & 17.71 & 2.76 & 7.27 & 32.25 & 0.12 & SK96 \\
$[+0015,-0029]$ & 37 & 280.0 & 3.54 & -3.07 & 5,222 & 3.03 & 0.14 & 14.19 & 2.58 & 11.47 & 37.13 & 0.11 & SK96 \\
$[-0019,+0013]$ & 23 & 15.0 & 4.14 & -3.23\tablenotemark{\dag} & 6,408 & 2.97 & 0.30 & 13.08 & 3.31 & 19.14 & 51.36 & 0.27 & SK96 \\
\enddata
\tablecomments{
See Table~\ref{tabhiiex} for column descriptions.}
\end{deluxetable}

\clearpage
\begin{deluxetable}{llccccccccc}
\tablecolumns{11} \tabletypesize{\scriptsize}
\tablewidth{0pt}
\tablecaption{Tully-Fisher Parameters\label{tabtf}}
\tablehead{
\colhead{GALAXY} & \colhead{T} & \colhead{$i$} & \colhead{$q$} & \colhead{$v_\odot$} & \colhead{$\vflat$} & \colhead{$I$} & \colhead{$\tauone$} & \colhead{$A_I$} & \colhead{$A^i_I$} & \colhead{$M_I$} \\
\colhead{} & \colhead{} & \colhead{$(\circ)$} & \colhead{\nodata} & \colhead{$(\kms)$} & \colhead{$(\kms)$} & \colhead{(mag)} & \colhead{} & \colhead{(mag)} & \colhead{(mag)} & \colhead{(mag)} \\
\colhead{(1)} & \colhead{(2)} & \colhead{(3)} & \colhead{(4)} & \colhead{(5)} & \colhead{(6)} & \colhead{(7)} & \colhead{(8)} & \colhead{(9)} & \colhead{(10)} & \colhead{(11)}
} \startdata
NGC3726 & Sc & 53 & 0.62 & 865.6 & $162\pm9$ & 9.51 & 0.019 & 0.03 & 0.18 & -21.98 \\
NGC3729 & Sab & 49 & 0.68 & 1059.8 & $151\pm11$ & 10.30 & 0.012 & 0.02 & 0.13 & -21.13 \\
NGC3917 & Scd & 79 & 0.24 & 964.6 & $135\pm3$ & 10.85 & 0.025 & 0.04 & 0.60 & -21.06 \\
NGC3949 & Sbc & 55 & 0.62 & 800.2 & $164\pm7$ & 10.28 & 0.024 & 0.04 & 0.18 & -21.22 \\
NGC3953 & Sbc & 62 & 0.5 & 1052.3 & $223\pm5$ & 9.02 & 0.034 & 0.05 & 0.36 & -22.67 \\
NGC4085 & Sc & 82 & 0.24 & 745.7 & $134\pm6$ & 11.28 & 0.020 & 0.03 & 0.59 & -20.62 \\
NGC4088 & Sbc & 69 & 0.37 & 756.7 & $173\pm14$ & 9.37 & 0.023 & 0.04 & 0.46 & -22.41 \\
NGC4100 & Sbc & 73 & 0.29 & 1074.4 & $164\pm13$ & 10.00 & 0.026 & 0.04 & 0.58 & -21.90 \\
NGC4102 & Sab & 56 & 0.56 & 846.3 & $178\pm11$ & 9.93 & 0.023 & 0.04 & 0.25 & -21.63 \\
NGC4138 & Sa & 53 & 0.63 & 893.8 & $147\pm12$ & 10.09 & 0.016 & 0.03 & 0.16 & -21.37 \\
UGC6399 & Sm & 75 & 0.28 & 791.5 & $88\pm5$ & 12.88 & 0.018 & 0.03 & 0.00 & -18.43 \\
UGC6446 & Sd & 51 & 0.62 & 644.3 & $82\pm4$ & 12.58 & 0.018 & 0.03 & 0.05 & -18.78 \\
UGC6667 & Scd & 89 & 0.12 & 973.2 & $86\pm3$ & 12.63 & 0.019 & 0.03 & 0.56 & -19.24 \\
UGC6917 & Sd & 56 & 0.54 & 910.7 & $104\pm4$ & 11.74 & 0.031 & 0.05 & 0.14 & -19.73 \\
UGC6983 & Scd & 49 & 0.66 & 1081.9 & $107\pm7$ & 11.91 & 0.031 & 0.05 & 0.09 & -19.51 \\
\cline{1-11} \\
Maffei2 & Sbc\tablenotemark{a} & 67\tablenotemark{b} & 0.421\tablenotemark{a} & -23\tablenotemark{b} & $170\pm 4$\tablenotemark{b} & 9.29\tablenotemark{a,*} & 2.017\tablenotemark{c,*} & 3.15\tablenotemark{c,*} & 0.37 & -21.87\tablenotemark{c,*} \\
\enddata
\tablecomments{
(1) Name of the galaxy.
(2) Morphological type from \citealt{ver01}.
(3) Adopted inclination angle from \citealt{ver01}.
(4) Ratio of the semi-minor to the semi-major axis from \citealt{vs01}.
(5) Heliocentric radial velocity from \citealt{ver01}.
(6) Mean rotational velocity of the flat part of the galaxy's rotation curve, corrected for inclination.
(7) Apparent total magnitude in $I$ from \citealt{ver01}.
(8) Optical depth of dust at 1~$\mu$m calculated from the
\citealt{sfd98} value of $E(B-V)$
(see \S~\ref{s_gasdust}).
(9) Galactic extinction correction in $I$
computed from \tauone\ using YES.
(10) Internal extinction correction in $I$ from YES
(see \S~\ref{s_dist_maf}).
(11) Absolute $I$ magnitude,
corrected for Galactic extinction and inclination.
K-corrections at the redshift of the UMa cluster
are negligible in the near-infrared.
The adopted distance modulus to the cluster is \muuma\ (see \S~\ref{s_zp}).
}
\tablenotetext{*}
{Uncertainties are given in Table~\ref{tabprop}.}
\tablenotetext{a}
{\citealt{bm99}}
\tablenotetext{b}
{\citealt{hth96}}
\tablenotetext{c}
{This study}
\end{deluxetable}

\clearpage
\begin{deluxetable}{lccccc}
\tablecolumns{5}
\tabletypesize{\scriptsize}
\tablewidth{0pt}
\tablecaption{Properties of Maffei 1, Maffei2 and IC 342\label{tabprop}}
\tablehead{
\colhead{Property} & \colhead{Symbol} & \colhead{Units} & \colhead{Maffei 1} & \colhead{Maffei 2} & \colhead{IC 342}
} \startdata
Apparent magnitudes\tablenotemark{a} & $B_T$ & mag & $13.47\pm 0.09$ & $14.77\pm 0.29$ & $9.37\pm 0.03$ \\
\nodata & $V_T$ & mag & $11.14\pm 0.06$ & $12.41\pm 0.08$ & $8.31\pm 0.03$ \\
\nodata & $I_T$ & mag & $8.06\pm 0.04$ & $9.29\pm 0.06$ & $6.68\pm 0.03$ \\
\cline{1-6} \\
Galactic optical depth at $\rm 1\mu m$ & $\tauone$ & \nodata & $1.691\pm 0.066$ & $2.017\pm 0.211$ & $0.677\pm 0.056$ \\
\cline{1-6} \\
Extinction normalized to $\tauone$ & $R^1_B$ & \nodata & $3.623$ & $3.654$ & $3.749$ \\
computed by YES & $R^1_V$ & \nodata & $2.765$ & $2.767$ & $2.831$ \\
\nodata & $R^1_I$ & \nodata & $1.571$ & $1.564$ & $1.588$ \\
\cline{1-6} \\
Galactic extinction & $A_B$ & mag & $6.128\pm 0.241$ & $7.371\pm 0.770$ & $2.537\pm 0.208$ \\
\nodata & $A_V$ & mag & $4.677\pm 0.184$ & $5.581\pm 0.583$ & $1.916\pm 0.157$ \\
\nodata & $A_I$ & mag & $2.658\pm 0.104$ & $3.155\pm 0.329$ & $1.075\pm 0.088$ \\
\cline{1-6} \\
Internal extinction & $A^i_B$ & mag & \nodata & 0.646 & 0.099 \\
(excess over face-on) & $A^i_V$ & mag & \nodata & 0.548 & 0.081 \\
\nodata & $A^i_I$ & mag & \nodata & 0.385 & 0.053 \\
\cline{1-6} \\
Apparent face-on magnitudes & $B_0$ & mag & $7.34\pm 0.26$ & $6.75\pm 0.82$ & $6.73\pm 0.21$ \\
corrected for Galactic extinction\tablenotemark{*} & $V_0$ & mag & $6.46\pm 0.19$ & $6.28\pm 0.59$ & $6.31\pm 0.16$ \\
\nodata & $I_0$ & mag & $5.40\pm 0.11$ & $5.75\pm 0.33$ & $5.55\pm 0.09$ \\
\cline{1-6} \\
Distance modulus & $\mu$ & mag & $27.28\pm 0.27$ & $27.62\pm 0.36$ & $27.41\pm 0.12$ \\
\cline{1-6} \\
Distance & $d$ & Mpc & $2.85\pm 0.36$ & $3.34\pm 0.56$ & $3.03\pm 0.17$ \\
\cline{1-6} \\
Absolute face-on magnitudes & $M_B$ & mag & $-19.93\pm 0.38$ & $-20.86\pm 0.90$ & $-20.68\pm 0.24$ \\
corrected for Galactic extinction & $M_V$ & mag & $-20.81\pm 0.33$ & $-21.34\pm 0.69$ & $-21.10\pm 0.20$ \\
 & $M_I$ & mag & $-21.87\pm 0.30$ & $-21.87\pm 0.49$ & $-21.86\pm 0.15$ \\
\enddata
\tablenotetext{*}
{Cosmological corrections at this redshift are negligible.}
\tablenotetext{a}
{\citealt{bm99}}
\end{deluxetable}

\clearpage
\figcaption[]{
\label{figtr}
Correlation between HII region temperature (T)
and galactocentric radius normalized to the
disk scale length ($r/r_0$) for HII regions in
the Sbc galaxy sample (filled circles)
and the Scd galaxy sample (open circles).
The rms deviation in temperature is 562 K for the
Sbc sample and 934 K for the Scd sample.
The linear fits represented by the solid lines
are given by Eq.~\ref{eqtsbc}
and Eq.~\ref{eqtscd}.
}
\figcaption[]{
\label{figdust}
Optical depth at 1~$\mu$m for extragalactic dust (\tauex)
versus the total column density of hydrogen (\nh) for HII regions
in Sbc galaxies (filled circles) and Scd galaxies (open circles).
The solid and open squares represent the total extinction observed
in the HII regions in Maffei~2 and IC~342, respectively.
The linear fit to the reference data (circles) is 
represented by the solid line and given by Eq.~\ref{eqdust}.
The rms scatter in \tauex\ is \rmstau.
The average vertical offset between the data points for
a heavily obscured galaxy and the reference HII regions
is the Galactic extinction of the heavily obscured galaxy.
}
\figcaption[]{
\label{figtf}
$I$-band TF relation for UMa spirals.
The solid line is the fit given by Eq.~\ref{eqtf}.
The rms scatter in $M_I$ is \rmstfi~mag.
The dashed line marks $\log{2\vflat}$ for Maffei~2.
}
\figcaption[]{
\label{figresid}
Difference between the total $\kp$ magnitude
from V01 and the extrapolated magnitude \kext\
from 2MASS as a function of the apparent central disk
surface brightness,
$\mu_0(\kp)$, from \citet{tul96}.
}
\clearpage
\plotone{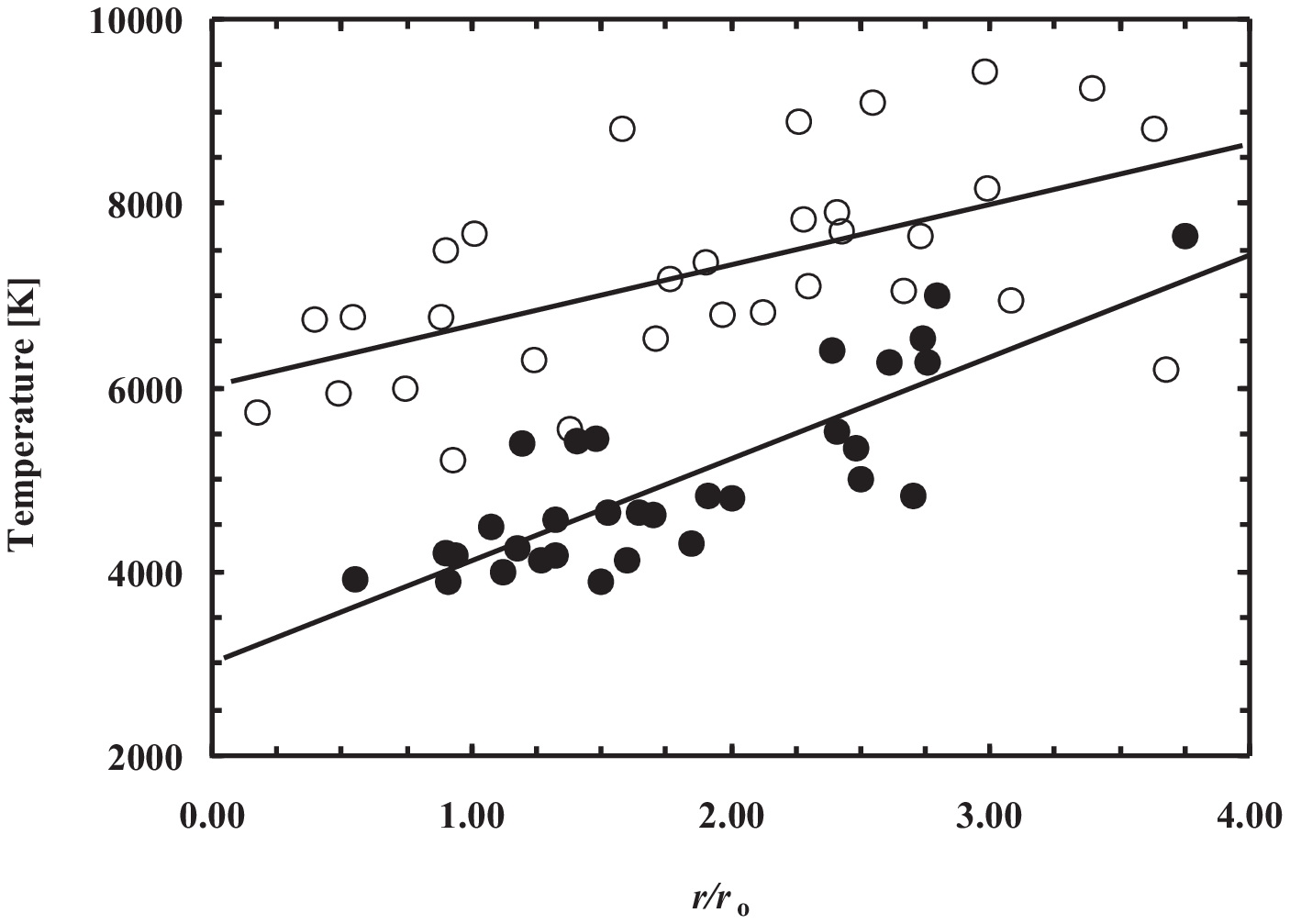}
\clearpage
\plotone{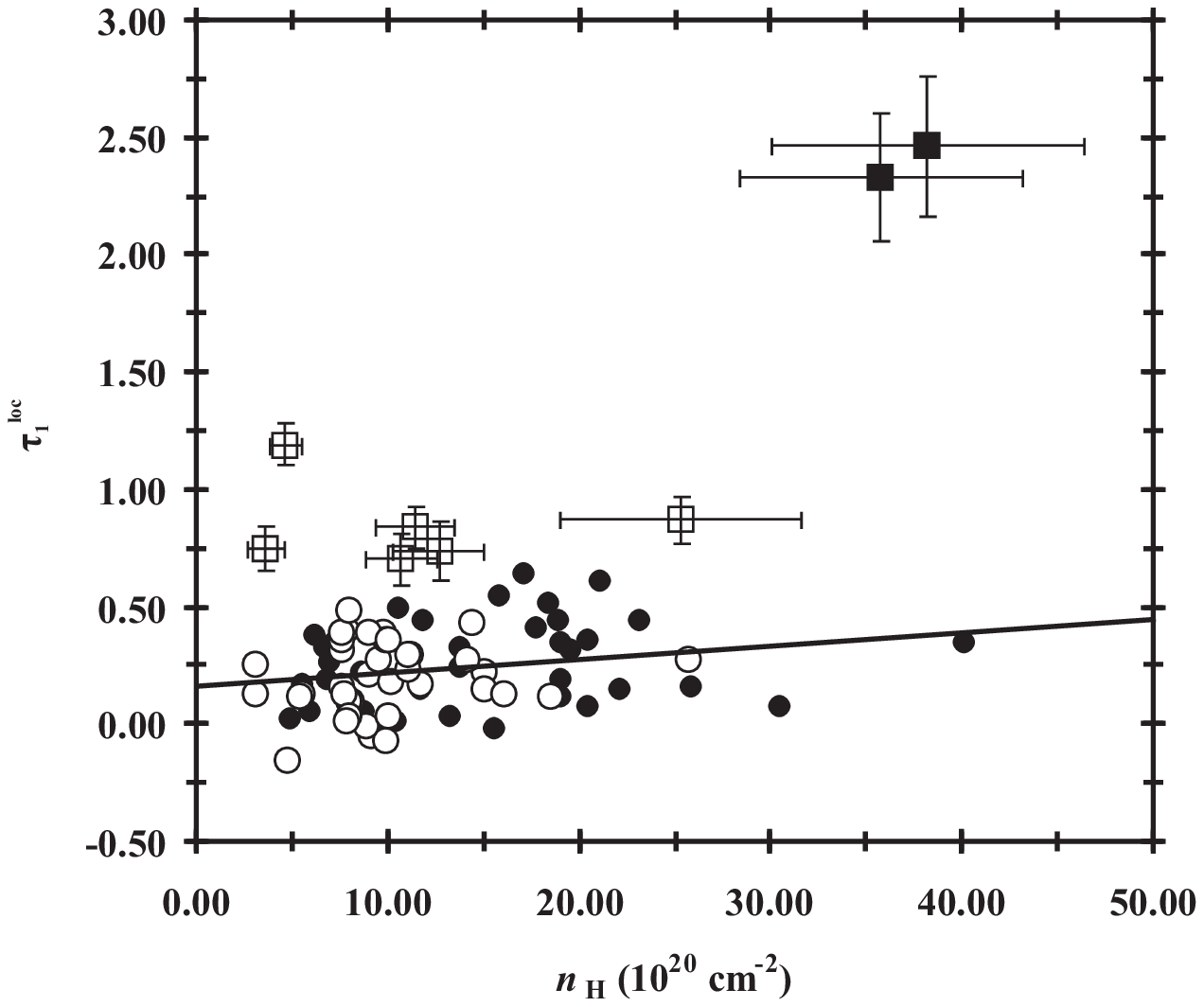}
\clearpage
\plotone{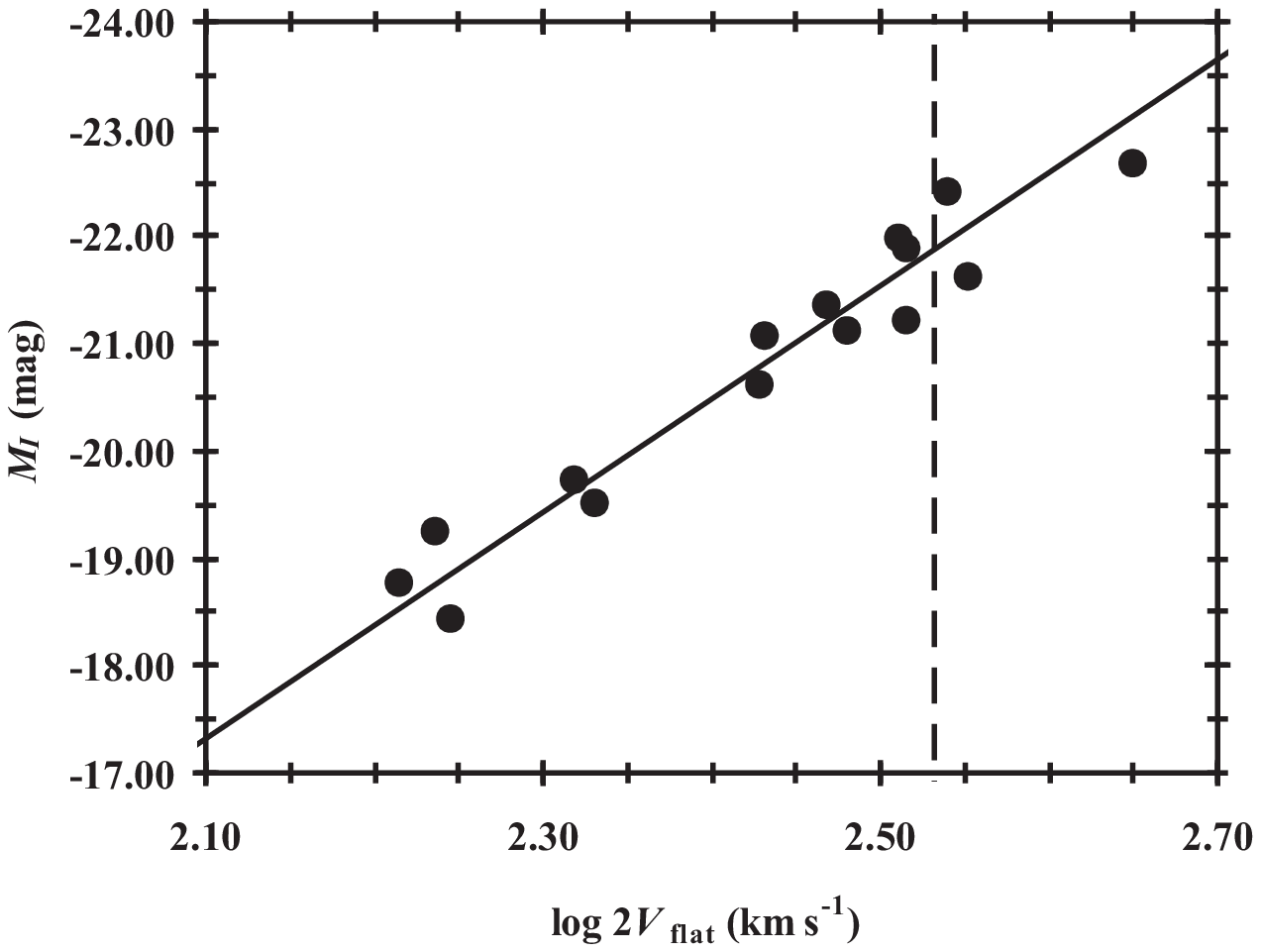}
\clearpage
\plotone{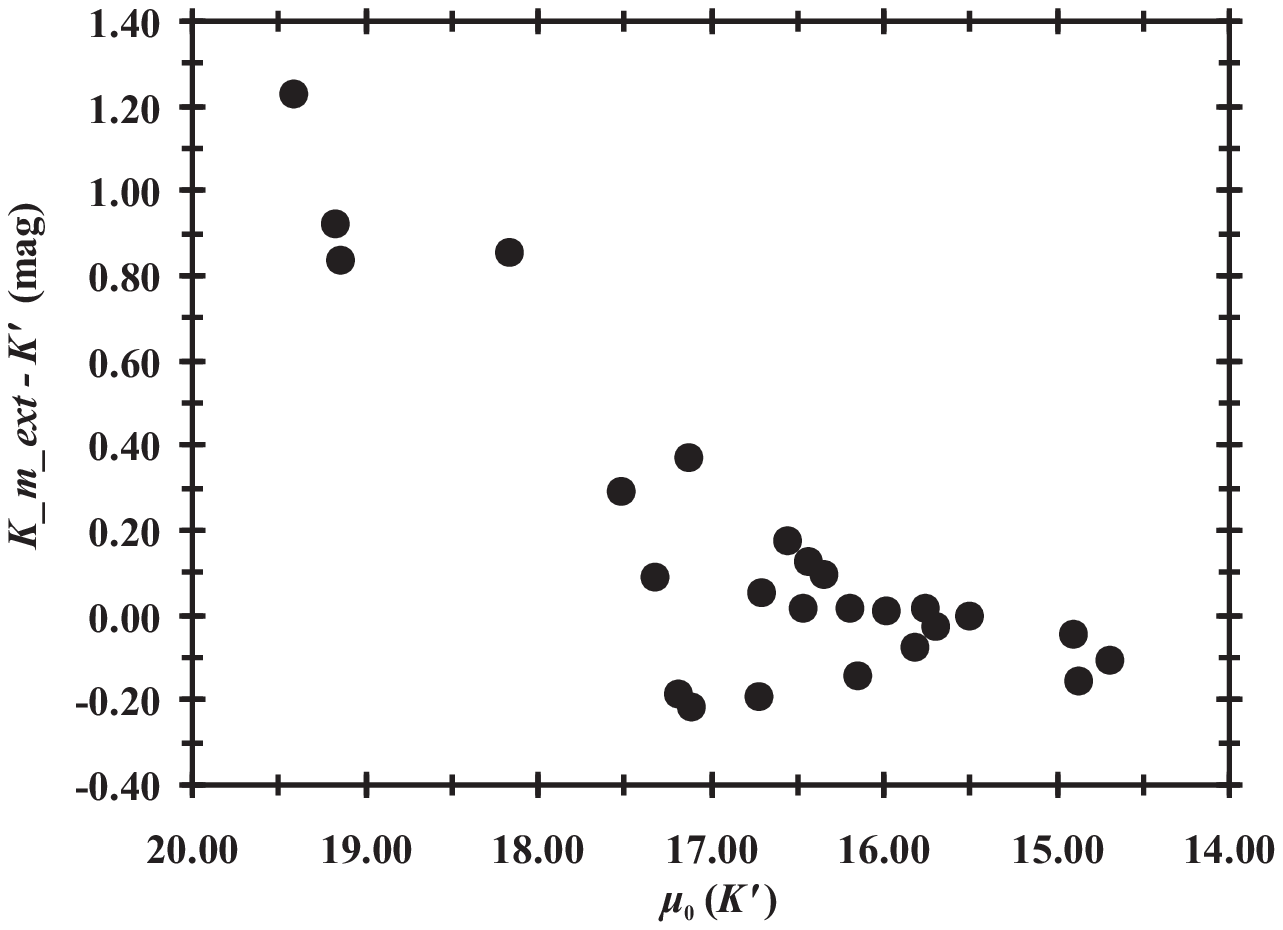}

\end{document}